\documentclass[11pt,a4paper,british]{article}
\pdfoutput=1
\usepackage{jinstpub}
\usepackage[T1]{fontenc}
\usepackage[utf8]{inputenc}
\usepackage{babel}
\usepackage{float}
\usepackage{textcomp}
\usepackage{amsmath}
\usepackage{amssymb}
\usepackage{upgreek}
\usepackage{graphicx}
\usepackage{physics}
\usepackage{eurosym}
\usepackage{tikz}
\usepackage{subcaption}
\usepackage[disable]{todonotes}
\usepackage[per-mode=symbol,separate-uncertainty=true, exponent-product=\cdot,multi-part-units=brackets,product-units=power]{siunitx}
\usepackage[super]{nth}

\usepackage[nottoc,notlof,notlot]{tocbibind}

\usepackage[section,above,below]{placeins}

\usepackage{xspace}
\usepackage{hyperref}

\title{An integrated general purpose SiPM based optical module with a high dynamic range}
\author[a,1]{T.~Bretz,\note{Corresponding author.}}
\author[b]{R.~Engel,}
\author[a]{T.~Hebbeker,}
\author[a]{J.~Kemp,}
\author[a]{L.~Middendorf,}
\author[a]{C.~Peters,}
\author[a]{J.~Schumacher,}
\author[a]{R.~\v{S}m{\'\i}da,}
\author[b]{and D.~Veberi\v{c}}
\affiliation[a]{III. Physikalisches Institut A, RWTH Aachen University \\ Otto-Blumenthal-Straße, 52074 Aachen, Germany}
\affiliation[b]{Institut für Kernphysik, Karlsruher Institut für Technologie \\ Hermann-von-Helmholtz-Platz 1, 76344 Eggenstein-Leopoldshafen, Germany}
\emailAdd{tbretz@physik.rwth-aachen.de}
\abstract{Silicon photomultipliers (SiPMs) are semiconductor-based light-sensors offering a high gain, a mechanically and optically robust design and high photon detection efficiency. Due to these characteristics, they started to replace conventional photomultiplier tubes in many applications in recent years. This paper presents an optical module based on SiPMs designed for the application in scintillators as well as lab measurements. The module hosts the SiPM bias voltage supply and three pre-amplifiers with different gain levels to exploit the full dynamic range of the SiPMs. Two SiPMs, read-out in parallel, are equipped with light guides to increase the sensitive area. The light guides are optimized for the read-out of wavelength shifting fibers as used in many plastic scintillator detectors. The optical and electrical performance of the module is characterized in detail in laboratory measurements. Prototypes have been installed and tested in a modified version of the Scintillator Surface Detector developed for AugerPrime, the upgrade of the Pierre Auger Observatory. The SiPM module is operated in the Argentinian Pampas and first data proves its usability in such harsh environments.}


\begin{document}
\maketitle
\flushbottom

\section{Introduction}
For many applications in astroparticle and high-energy physics, highly durable and low-power optoelectronic modules are of major importance. These applications include detector systems using plastic scintillators, read-out by wavelength shifting fibers, and also characterization equipment for the use in laboratories. For many of these applications, a high dynamic range as well as a precise calibration of the light flux is crucial. A module addressing these requirements based on silicon photomultipliers was developed.

The overall design of the module was optimized for replacing conventional photomultiplier tubes (PMTs) especially in scintillator detectors. Typical PMTs used in astroparticle physics experiments have peak quantum efficiencies of \SIrange{20}{30}{\percent} and a gain in the order of $10^4$ to $10^6$. Their linear dynamic range reaches up to some \num{10000} photons within \SI{\sim100}{\ns} depending on the gain~\cite{HamamatsuR9420}. With the developed silicon photomultiplier based optical module, a performance is achieved which is at least as good as that of conventional PMTs.

\subsection{Silicon photomultipliers}\label{sec:SiPMs}

Silicon photomultipliers (SiPMs) are cell-structured, photo-sensitive semiconductors superseding conventional photomultiplier tubes in many applications in high-energy and astroparticle physics where the demand of high durability and performance is more significant than large collection areas. Each cell is an avalanche photodiode operated in Geiger mode (G-APD). Single SiPM devices are commercially available in sizes from \SI{1}{\square\mm} up to \SI{36}{\square\mm} and cell pitches as large as \SI{100}{\um} down to \SI{10}{\um}~\cite{Hamamatsu2015TechInfo, Sensl2015CSeries}. They consist of one hundred up to some ten thousand cells.

To operate SiPMs, a bias voltage $V_\text{b}$ larger than the breakdown voltage $V_\text{br}$ of the photodiode is applied. Basic SiPM characteristics such as gain, photon detection efficiency (PDE) and crosstalk depend on the overvoltage $V_\text{ov}=V_\text{b}-V_\text{br}$. The breakdown voltage depends on temperature. Thus, the bias voltage has to be adjusted according to the temperature to achieve a temperature-independent operation. SiPMs are biased with moderate operating voltages below \SI{100}{\volt}~\cite{HamamatsuS133606025PE, Sensl2015CSeries}.

When a SiPM cell is hit by a photon, its charge is released in a characteristic pulse, referred to as photoelectron equivalent (p.e.). After the discharge, the gain and PDE of the cell are reduced until the cell is recharged again. This effect together with the limited number of cells limits the dynamic range and alters the response especially for bright long-lasting pulses where cells get hit multiple times. A correction can be applied to reconstruct the incident light pulse~\cite{SiPMDynRangeICRC}. A homogeneous light distribution on the SiPM is preferable to distribute the light over as many cells as possible.

A typical SiPM has three sources of noise. The main contribution is dark noise randomly induced by thermal excitations in a cell initiating an avalanche. \emph{Optical crosstalk} originates from photons that are produced during a cell breakdown and trigger a neighbouring cell. In addition, \emph{after pulsing} occurs from electrons or holes trapped in impurities of the SiPM lattice structure. They are released with a delay and induce a new avalanche. For recent devices the crosstalk probability is below \SI{10}{\percent} and the after pulsing probability below \SI{1}{\percent}~\cite{Hamamatsu2015TechInfo, Sensl2015CSeries}.

SiPMs offer single-photon resolution allowing for a precise calibration of the incident light flux and have similar or higher photon detection efficiencies than conventional photomultiplier tubes (PMTs)~\cite{HamamatsuS133606025PE,Sensl2015CSeries}. In addition, they are tolerant against the exposure with bright light~\cite{FACTICRC2013}.

Due to these advantages, SiPMs are replacing conventional PMTs in many applications. They have been operated in the First G-APD Cherenkov Telescope (FACT)~\cite{FACT} for several years now with great success and new small size telescopes are being developed for the Cherenkov Telescope Array (CTA)~\cite{CTASST}. Because of their small size, SiPMs also allow to develop light-weight refractive telescopes for the observation of extensive air showers~\cite{FAMOUS}. SiPMs are tested for the scintillator detector upgrade of the IceTop array~\cite{IceTopUpgrade} and are used in high-energy particle physics experiments like the Hadronic Calorimeter of the Compact Muon Solenoid (CMS) detector~\cite{CMSHODesign,CMSHOOperation} where they have been successfully operated for several years.

\section{Overview}

\begin{figure}
	\includegraphics[width = .49\textwidth, keepaspectratio]{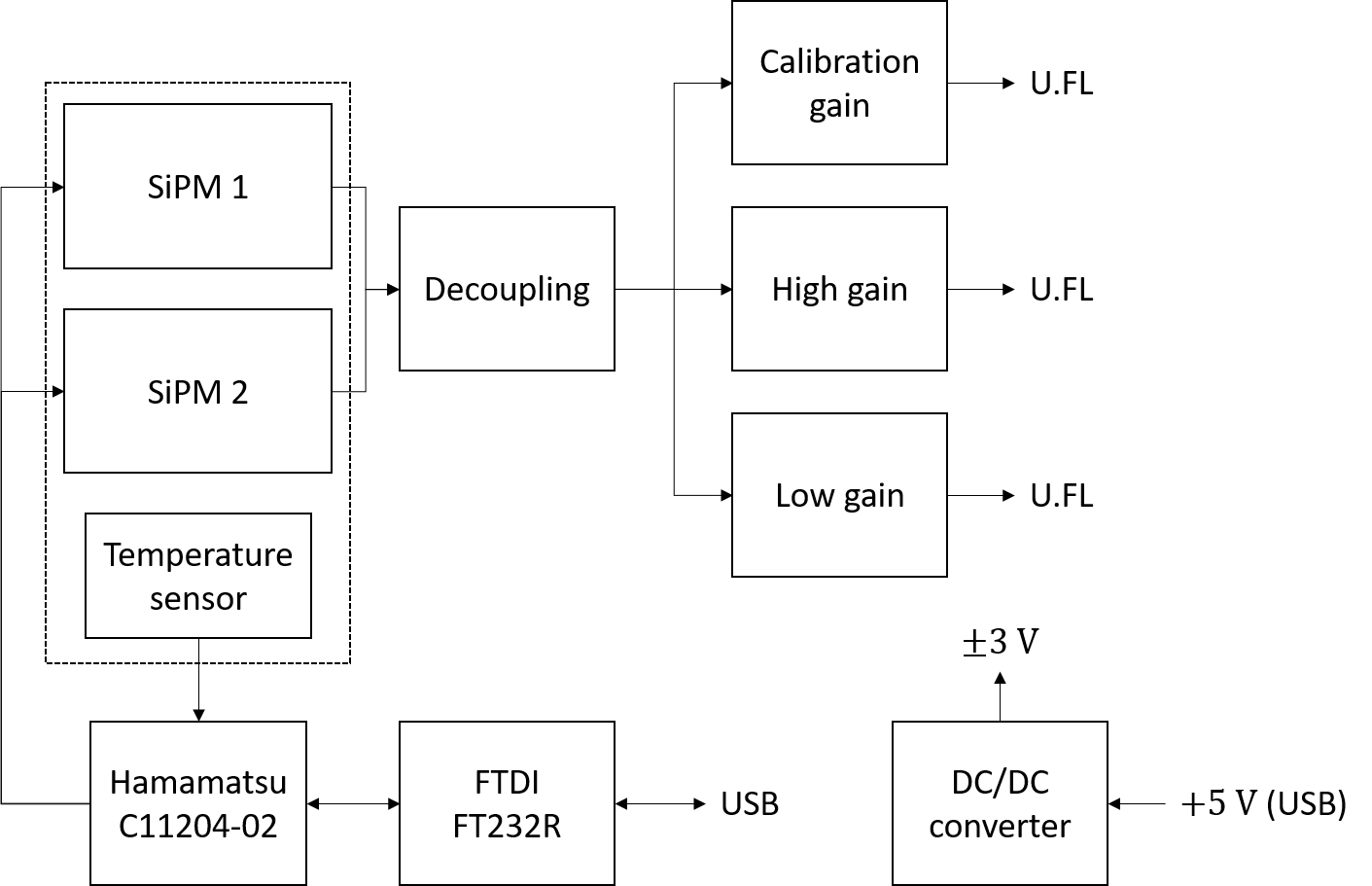}
    \hfill 
	\includegraphics[width = .49\textwidth, keepaspectratio,trim={210 0 80 0},clip]{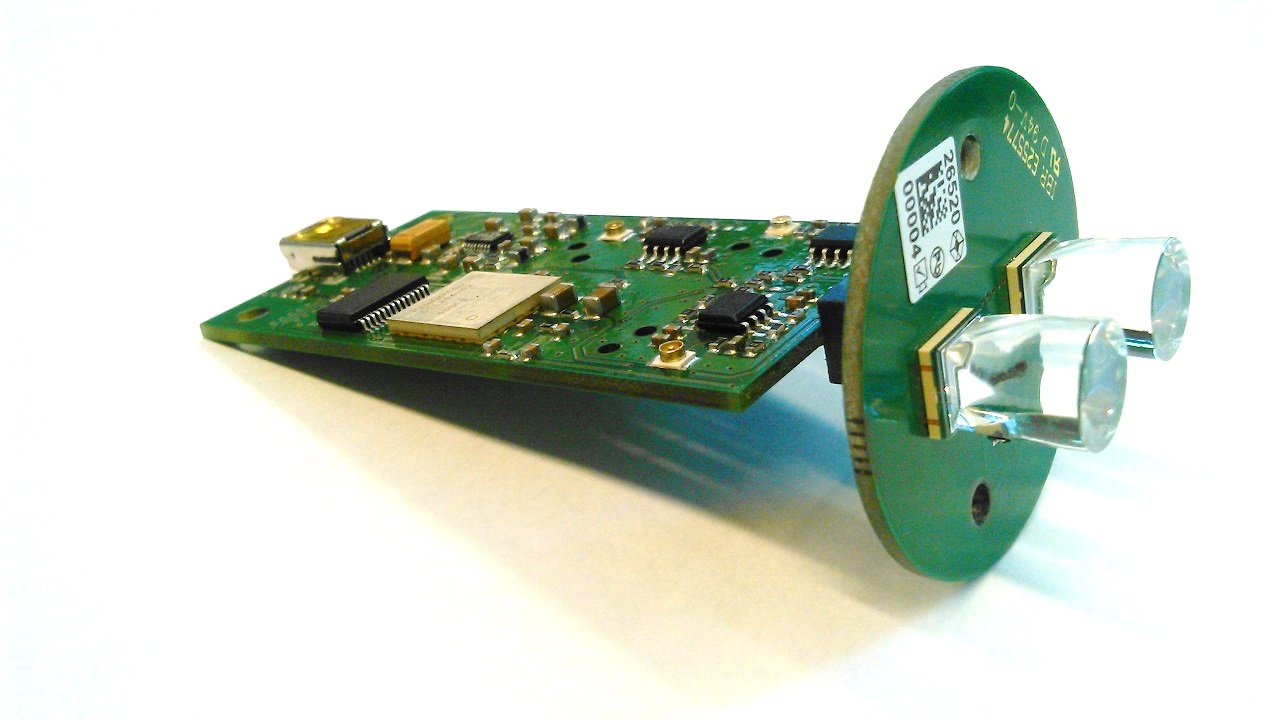}
\caption{\label{fig:overview} \emph{Left:} The schematic overview of the SiPM module. The Hamamatsu IC C11204-02 generates the regulated bias voltage for the two silicon photomultipliers (SiPMs) and is controlled and powered via USB. The signal of the two SiPMs is decoupled, summed up and amplified by three independent gain channels. The system is supplied via the +5\,V USB bus and draws less than 40\,mA, see text for details. \emph{Right:} A picture of the two printed circuit boards: The circular SiPM PCB on the right, hosting SiPMs, light guides and identification number; and the rectangular PCB on the left containing slow control, pre-amplifiers, and connectors.}
\end{figure}

In this application, two rectangular \SI{6 x 6}{\mm} surface-mounted SiPMs with \SI{25}{\um} cell pitch of type Hamamatsu S13360-6025PE were chosen~\cite{HamamatsuS133606025PE}. They offer a wide dynamic range of several \num{10000} photons with a total of \num{2 x 57600} cells and splendid optical performances with \SI{25}{\percent} peak photon detection efficiency at only \SI{1}{\percent} optical crosstalk. As the gain of the device is only $7\times10^5$ and the number of cells is enormous, care must be taken in developing pre-amplifiers addressing the whole dynamic range with a sufficient signal-to-noise ratio, to achieve reasonable resolution from single-p.e.\ up to a breakdown of all cells (see sec.\ \ref{sec:Electronics}). Choice of these SiPMs allows to reach similar optical performance as with conventional PMTs.

Since the dark count rate of SiPMs scales linearly with their surface area, the surface area is kept as low as possible and only two SiPMs are used. A larger sensitive area would be advantageous when the module is used in detectors designed for conventional PMTs with a large cathode area. The total sensitive area of \SI{72}{\square\mm} of both SiPMs is still small compared to conventional PMTs. Solid PMMA (Polymethyl methacrylate) light guides optimized for the application with wavelength shifting fibers have been developed to increase the sensitive area.

A schematic overview of the module is shown in figure~\ref{fig:overview} (left). The module consists of two printed circuit boards (PCBs). The main PCB contains the DC/DC converters, pre-amplifiers, and connectors, whereas the circular-shaped PCB holds the two SiPMs, the solid light guides and the analog temperature sensor. This mechanical design allows for easy replacement of conventional PMTs with only small modifications of the detector.

The module is supplied and controlled via USB and the analog output of the SiPMs is amplified by three independent gain stages providing the full dynamic range over more than five orders of magnitude (1\,p.e.\ up to a breakdown of all cells at the same time corresponding to \num{115200}\,p.e.).

Figure~\ref{fig:overview} (right) contains a picture of one fully assembled module. The individual components on the SiPM PCB on the right and the main PCB on the left are described in detail in the following sections.

\section{Optics}
	\subsection{Design goals}\label{sec:DesignGoals}
	The particular design of the solid PMMA light guides mainly depends on the sensitive area to be achieved and the angular distribution of the incoming light. For the application in scintillator detectors, the light guides are optimized for reading-out of a bundle of wavelength shifting fibers. Fiber bundles are easiest arranged in a round shape. Thus, the light guides transform from a round entry-window to a square exit-window with the size of the SiPM.
    
    To avoid inhomogeneities over the detector, the transmission efficiency of the light guides should be independent of the specific position of the fiber in the bundle. As described in section~\ref{sec:SiPMs} all SiPM cells should have equal hit probability and the spatial light distribution on the SiPM should be uniform to achieve a high dynamic range.
    
    A modified version of the Scintillator Surface Detector (SSD)~\cite{ICRCSSD} designed for an upgrade of the Pierre Auger Observatory~\cite{PAO,AugerPrime} was used to test the optical module. It hosts 48 scintillator bars with a size of \SI{160 x 5 x 1}{\cm} arranged in two detector halves with the optical module in between. In total, 48 wavelength shifting fibers with a diameter of \SI{1}{\mm} each are put through the scintillator bars to guide the produced light onto the optical module. The fibers used in this application have their peak emission at \SI{480}{\nm} (Kuraray Y11,~\cite{KurarayY11}). To match the geometry of the SiPM module, the 96 fiber ends of the detector are arranged in two fiber bundles of 48 fibers with a diameter of \SI{8}{\mm} (or an area of \SI{\sim50}{\square\mm}) each. This is significantly larger than the area of \SI{36}{\square\mm} of one SiPM. The fiber bundles are fixed with Saint-Gobain BC-600 optical cement~\cite{BC600} and protected by a \SI{7.5}{\mm} thick PMMA window with a diameter of \SI{9}{\mm} surrounded by a reflective aluminium ring. 
    \begin{figure}\centering
    	\includegraphics[width=0.7\textwidth]{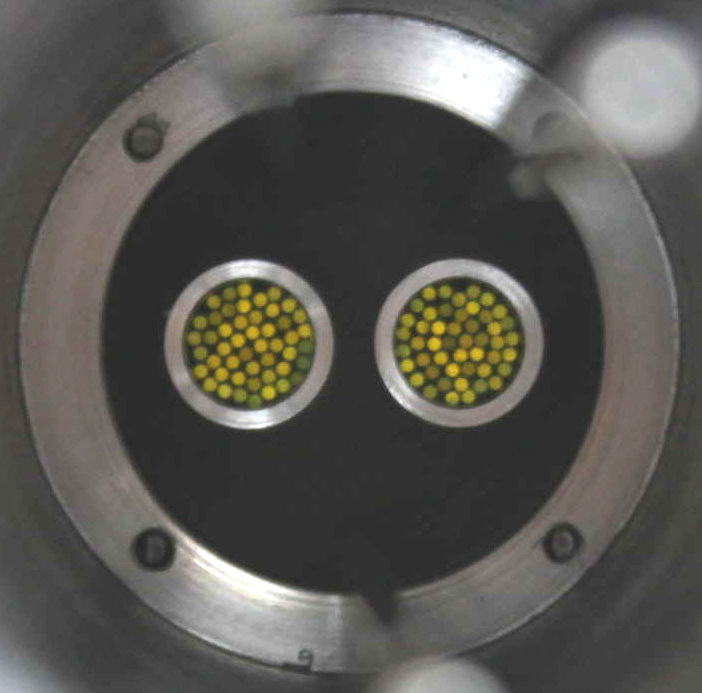}
        \caption{A front view of the holding structure for the two fiber bundles as it is already installed inside the detector. Each fiber bundle is surrounded by an aluminium ring. The three blurred pins (two at the top, one at the bottom) allow to precisely position the SiPM module in front of the fiber bundles.}
        \label{fig:SiPMCookiePhoto}
    \end{figure}
    A front view of this holding structure is shown in figure~\ref{fig:SiPMCookiePhoto}.
    
    The design for the light guides presented in the following sections is optimized for this application.
	
	\subsection{Light guides}
	The light guides are made of PMMA as it ensures a good optical transmission efficiency for green light emitted by wavelength shifting fibers and also allows for easy production through milling or casting. A polished surface ensures total reflection.
	To build a design fulfilling the mentioned constraints, an intersection of a round and a squared geometrical body will be used. The simplest realization of the round part is given by a cylinder or a truncated cone. The square part can consist of a truncated pyramid with or without parabolic surfaces.
    
    The parabolic surfaces follow the tilted parabolic shape of a Winston cone~\cite{WinstonCone} (referred to as \emph{Winston pyramid} in this document). An ideal Winston cone is rotational symmetric and allows for perfect compression of light for incident angles below a cut-off angle $\theta_{\text{max}}$.
    
    \begin{figure}\centering
    	\begin{minipage}{0.3\textwidth}
    		\includegraphics[width=\textwidth]{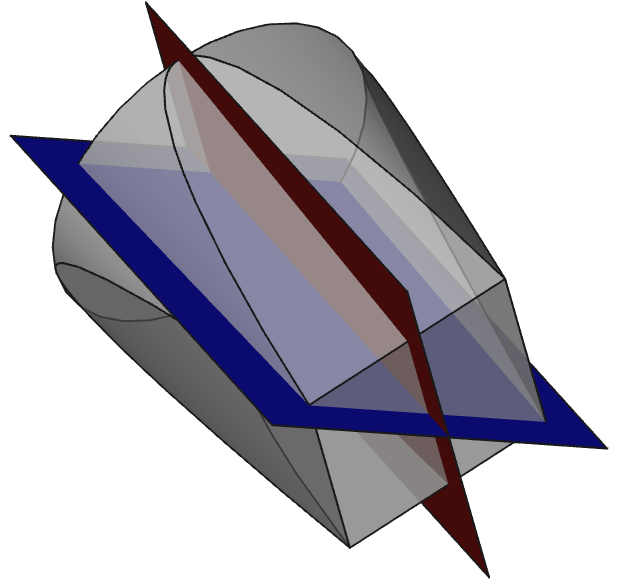}
    	\end{minipage}
        \hfill
        \begin{minipage}{0.3\textwidth}
    		\includegraphics[width=\textwidth]{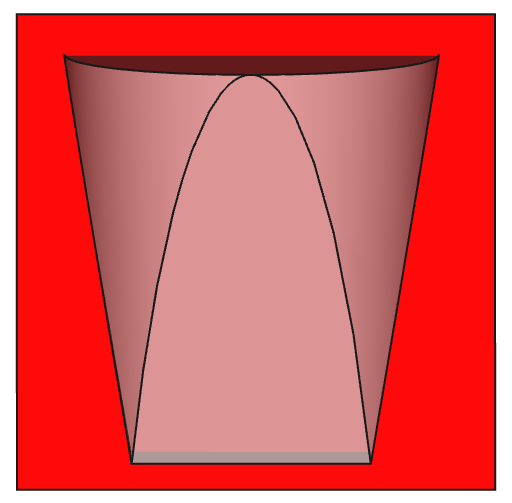}
    	\end{minipage}
        \hfill
        \begin{minipage}{0.3\textwidth}
    		\includegraphics[width=\textwidth]{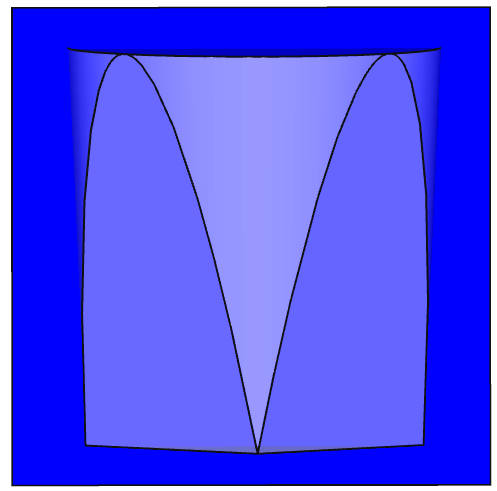}
    	\end{minipage}
        \begin{minipage}{0.3\textwidth}
    		\includegraphics[width=\textwidth]{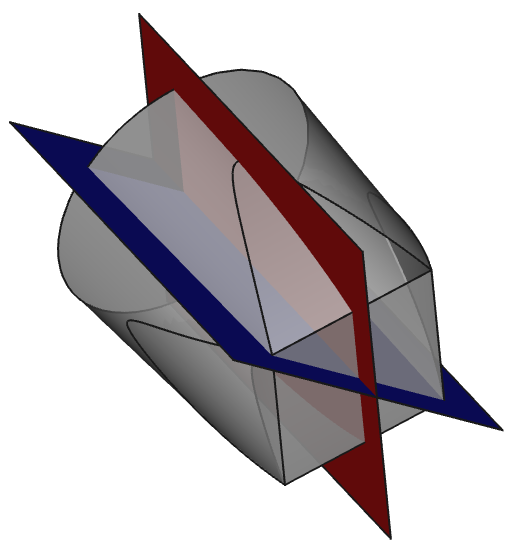}
    	\end{minipage}
        \hfill
        \begin{minipage}{0.3\textwidth}
    		\includegraphics[width=\textwidth]{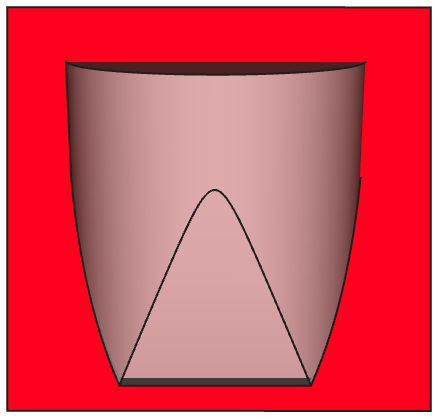}
    	\end{minipage}
        \hfill
        \begin{minipage}{0.3\textwidth}
    		\includegraphics[width=\textwidth]{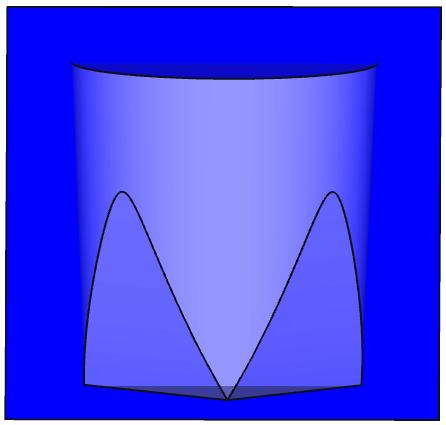}
    	\end{minipage}
        \caption{Examples of the resulting light guide geometries. Both shown light guides will be studied in detail later. \emph{Top row:} A light guide consisting of a regular pyramid and a cone with a length of 1\,cm shown from three different perspectives. All surface contours follow straight lines. \emph{Bottom row:} A light guide consisting of a Winston pyramid and a cone with a cut-off angle of \ang{39}. Here, the Winston pyramid causes parabolically shaped surface contours.}
        \label{fig:LightGuideGeometry}
    \end{figure}
    A sketch of the resulting light guide geometries is shown in figure~\ref{fig:LightGuideGeometry}. All the geometries can be varied in length or cut-off angle, respectively.
    
	\subsection{Simulation}
\begin{figure}\centering
		\includegraphics[width=\textwidth]{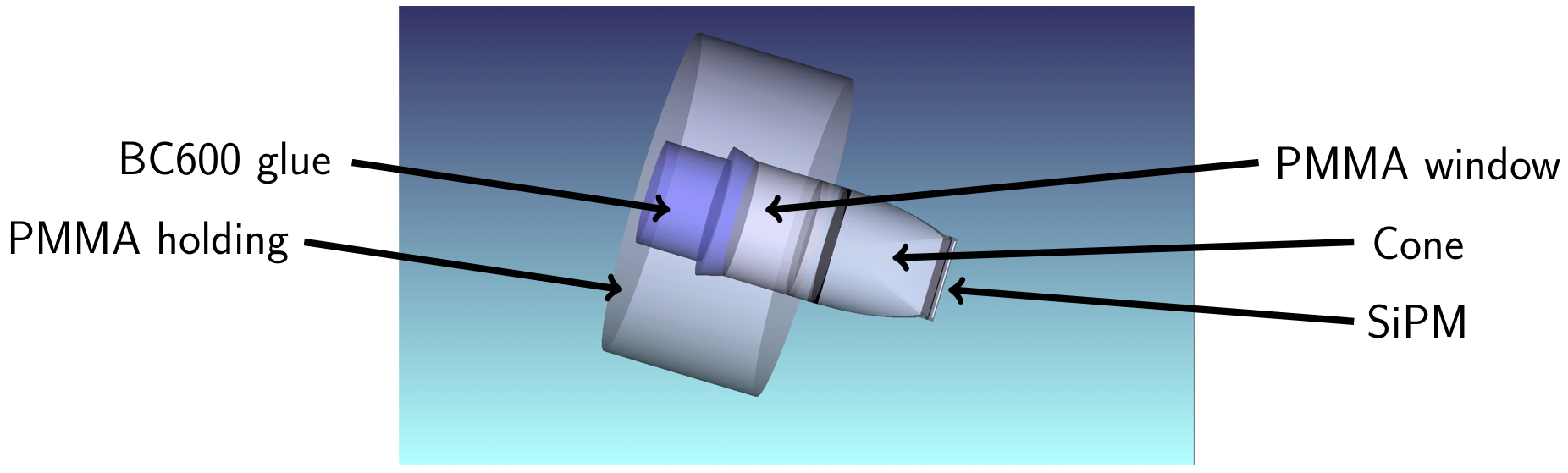}
		\caption{The basic design for the simulation. All optical components including the Saint-Gobain BC-600 optical cement, the PMMA holding structure and the reflective aluminium ring are implemented. Only one fiber bundle is simulated because the distance between the two fiber bundles is large enough to be optically independent.}
		\label{fig:Sim_Design}
	\end{figure}

	The simulations are implemented in Zemax Optical Studio 12. The basic design of the simulation is shown in figure~\ref{fig:Sim_Design}.
    \begin{figure}\centering
    	\includegraphics[]{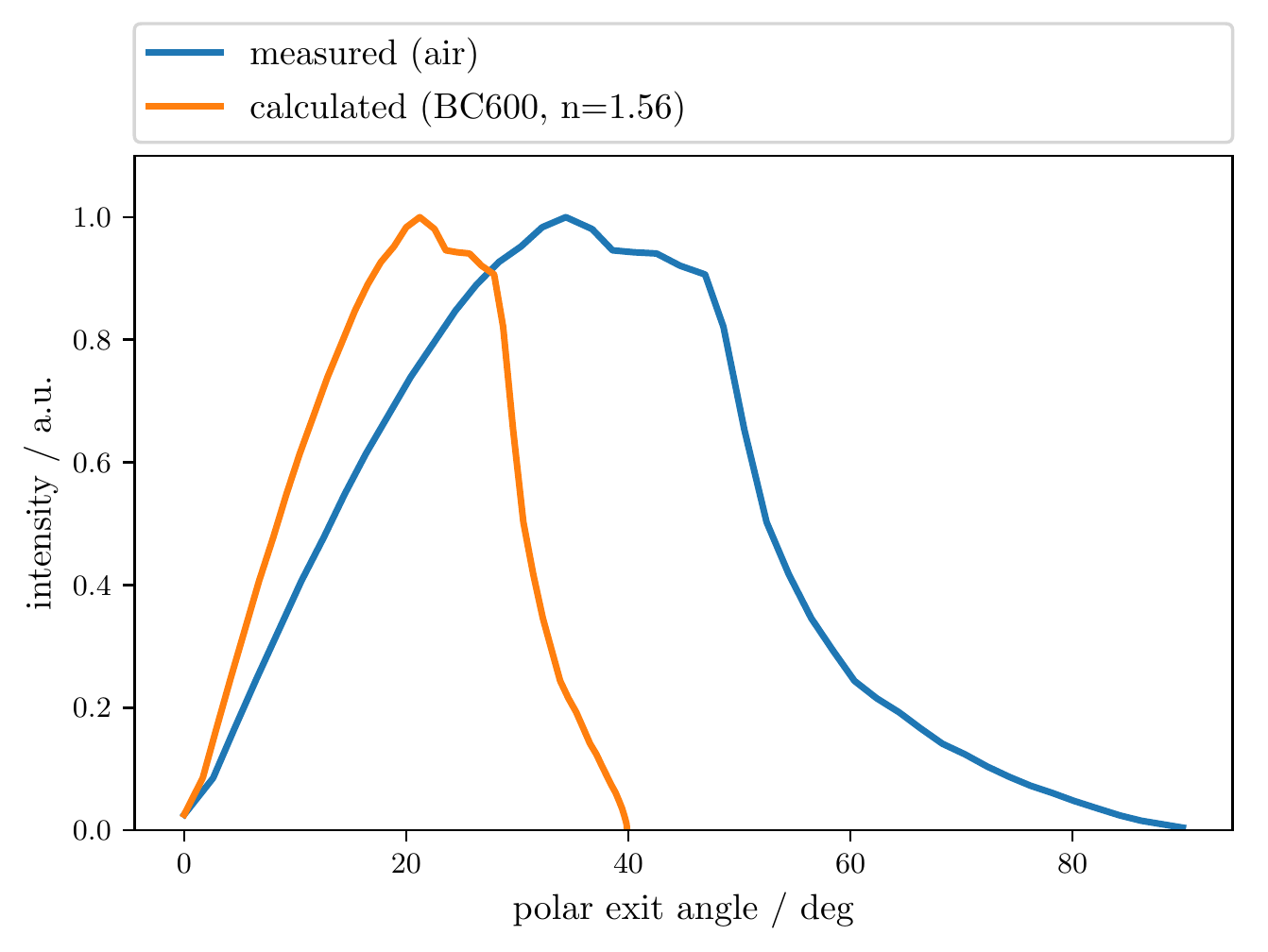}
		\caption{The angular distribution of the light output of the used optical fibers. The blue curve shows the measurement in air and the orange curve the corresponding calculation using Snell's law for the transition from air to Saint-Gobain BC-600 glue~\cite{BC600}.}
		\label{fig:FiberLightOutput}
	\end{figure}
	All optical components of the structure described in section~\ref{sec:DesignGoals} are simulated except for the fibers. The angular distribution of the fiber light output has been measured in the lab (see fig.\ \ref{fig:FiberLightOutput}~\cite{ThesisNieswand}). Each fiber is represented by a round light source with a diameter of \SI{1}{\mm} emitting photons according to this angular distribution at each point. 
    
    To simplify the simulation, only 37 fibers (hexapolar structure with four rings, already implemented in Zemax) instead of 48 are used. They are spaced such that they occupy all the space available in the \SI{8}{\mm} diameter tube.
	
    
    The PDE of the SiPM is angular dependent due to reflections at the coating and the silicon surface. This is taken into account by simulating all optical properties of the SiPM including those of the coating and the silicon surface. Only photons entering the silicon are considered to be detected.
    
	\subsubsection{The geometries of the light guides}
	Four different cone geometries are considered for the simulation:
	\begin{enumerate}
		\item Quadratic part: Winston pyramid; round part: cylinder
		\item Quadratic part: Winston pyramid; round part: cone
		\item Quadratic part: regular pyramid; round part: cylinder
        \item Quadratic part: regular pyramid; round part: cone
	\end{enumerate}
	The size of the entrance window of a Winston pyramid is fixed if its exit-window size and its cut-off angle is fixed. In case the entrance window is too large, the light guide is truncated such that the remaining stump fits the size of the fiber bundle. Truncating the Winston pyramid does not preserve the cut-off angle. Thus, smaller nominal cut-off angles might have better performance than larger cut-off angles. The maximum possible cut-off angle for an ideal Winston cone with an entrance window radius $R=4.7\,$mm and an exit-window radius $r=3\,$mm is
	\begin{align}
		\theta_\text{max}=\arcsin\left(\frac{r}{R}\right)=39.7^\circ
	\end{align}
    The angle equals the maximum angle of the light exiting the fiber (fig.\ \ref{fig:FiberLightOutput}). For an ideal Winston cone, all light is guided onto the SiPM.
    \par
	In the simulation, different cut-off angles $\theta_\text{max}$ and lengths are tested with respect to the transmission efficiency and the homogeneity on the SiPM. The entrance radius for all simulations is \SI{4.7}{\mm} to allow for some misplacement with respect to the holding structure with a radius of \SI{4.5}{\mm}. In addition, the efficiency for a single fiber at different positions in the fiber bundle is studied.
	
	\subsubsection{Light guide efficiencies}
\begin{figure}\centering
    	\includegraphics[]{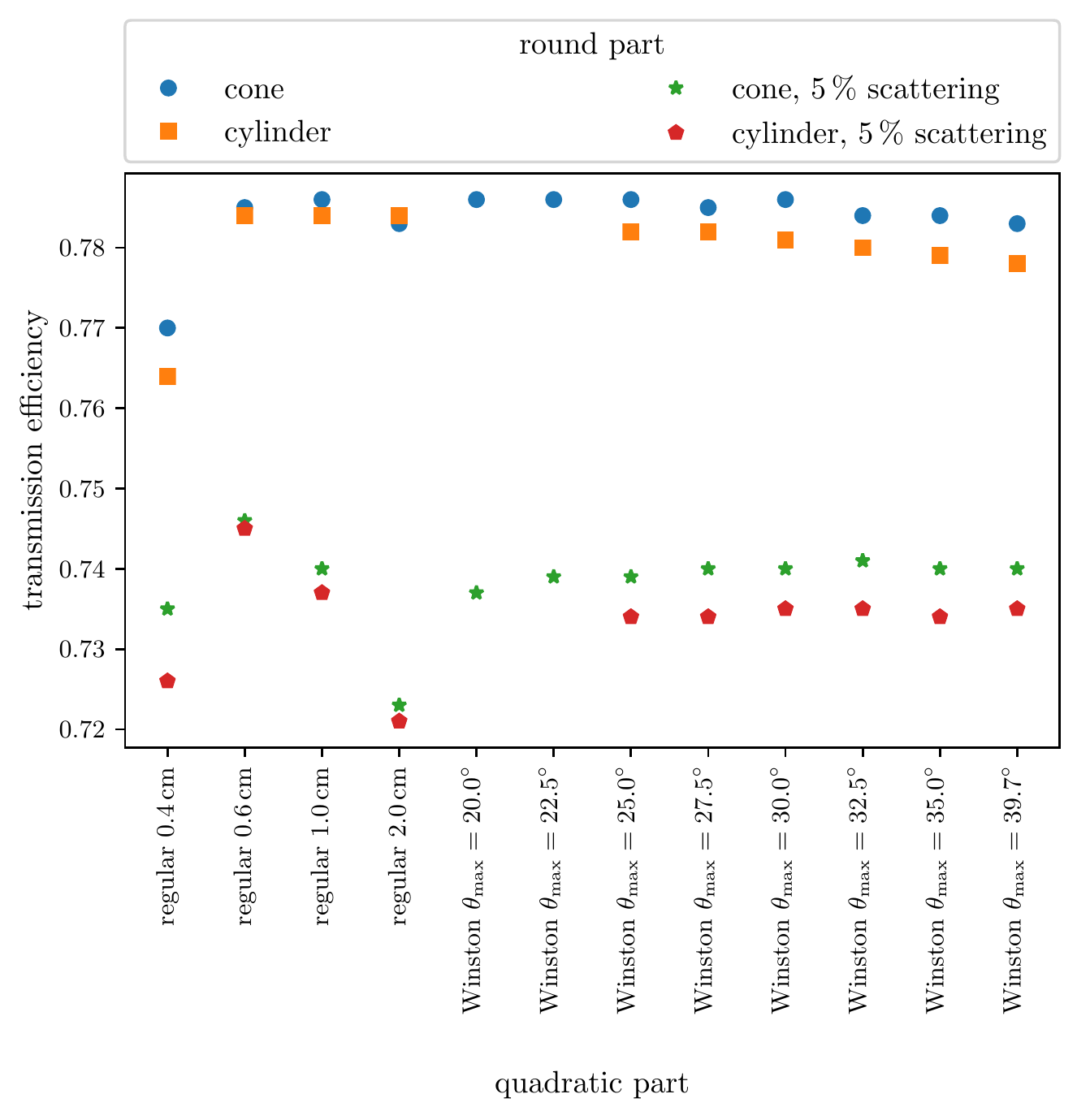}
		\caption{The total light guide transmission efficiency for 37 fibers and light guides with an entry-window radius of 4.7\,mm to allow a slight misalignment. The simulation is performed for light guides with an ideal surface and for light guides with \SI{5}{\percent} Lambertian scattering.}
         \label{fig:Efficiency}
    \end{figure}

	The total transmission efficiency with 37 fibers for different simulated light guide geometries is shown in figure~\ref{fig:Efficiency}.
    For ideal surfaces (blue circles, orange squares), the relative differences between the transmission efficiencies of the light guides are only at the percent level. Light guides with a length of \SI{0.6}{\cm} or more show the same performance. Only shorter light guides have a significantly worse performance. This is expected as for these light guides, the angle between the incoming rays and the surface can already exceed the maximum angle for total reflections. Thus, some light is transmitted out of the light guide. Winston geometries show similar performance as regular pyramid configurations.
	
    Differences are small due to the simulation of perfectly reflecting surfaces. Rays can be reflected multiple times inside a light guide without getting transmitted out of it. Most light is lost at optical transitions.
	
	To understand the effect of non-specular reflections, a Lambertian scattering of \SI{5}{\percent} of the incident rays at the surface of the light guides was included in the simulation. Compared to no scattering, the transmission efficiency is reduced by roughly \SI{5}{\percent}. The transmission efficiency of the light guides increases with its cut-off angle. For large cut-off angles, light guides are shorter and thus fewer reflections at their surface occur. The same argument holds for light guides with regular pyramid surfaces.
	
	In this simulation, shorter light guides or larger cut-off angles, respectively, are preferable. As expected, the light guides with a cone as the round part perform better than those with a simple cylinder. Thus, all following simulations are performed with a cone.
    
    The photon detection efficiency of the SiPM taken from the datasheet (\SI{25}{\percent} at \SI{480}{\nm}~\cite{HamamatsuS133606025PE}) is measured in air for vertically incident light and includes the reflections at the optical transitions of the SiPM coating and silicon of \SI{\sim75}{\percent} in total. Taking this effect into account, the probability for a photon that entered the surface to create an avalanche is \SI{\sim33}{\percent}. Using the simulation, a transmission efficiency from the fibers through the light guide into the silicon of \SI{74}{\percent} is determined (c.f.~fig.~\ref{fig:Efficiency}). The photon detection efficiency of the complete optical system including the light guide is thus given by $0.74\times0.33=24\,\%$. This is better by \SIrange{20}{60}{\percent} compared to a quantum efficiency of \SIrange{15}{20}{\percent} at \SI{480}{\nm} of conventional PMTs~\cite{HamamatsuR9420}.
    
    Local imperfections in the shape might result from the polishing of the light guide. As it only affects small areas on the surface of the light guide, its effect is negligible compared to scattering. Hence, it is not studied here.\\
    
	The following sections focus on comparing the properties of two light guides:
    \begin{enumerate} 
    	\item A regular pyramid with 1\,cm length intersected with a cone: It is significantly longer than the badly performing 0.4\,cm long light guide and, thus allows for some safety margin due to imperfections in the simulation. It is not yet too long and thus has a good transmission efficiency. In the following, it is named \emph{\mbox{regular-shaped light guide}}.
        \item The light guide with a Winston pyramid with $\theta_{\text{max}}=39^\circ$ intersected with a cone: This is the shortest possible light guide with a Winston pyramid and thus promises the best performance. It is named \emph{\mbox{Winston-shaped light guide}} hereafter.
	\end{enumerate}
    Both light guides were already presented in figure~\ref{fig:LightGuideGeometry}.
	
	\subsubsection{Fiber position}
\begin{figure}
		\begin{subfigure}[t]{0.5\textwidth}\centering
			\includegraphics[]{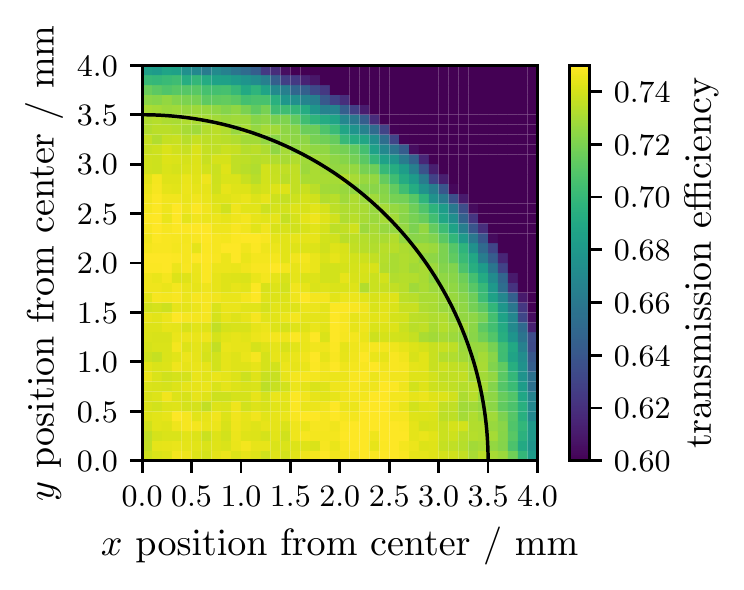}
			\caption{The regular-shaped light guide.}
			\label{fig:FiberPosition_Normal}
		\end{subfigure}
        \hfill
		\begin{subfigure}[t]{0.5\textwidth}\centering
			\includegraphics[]{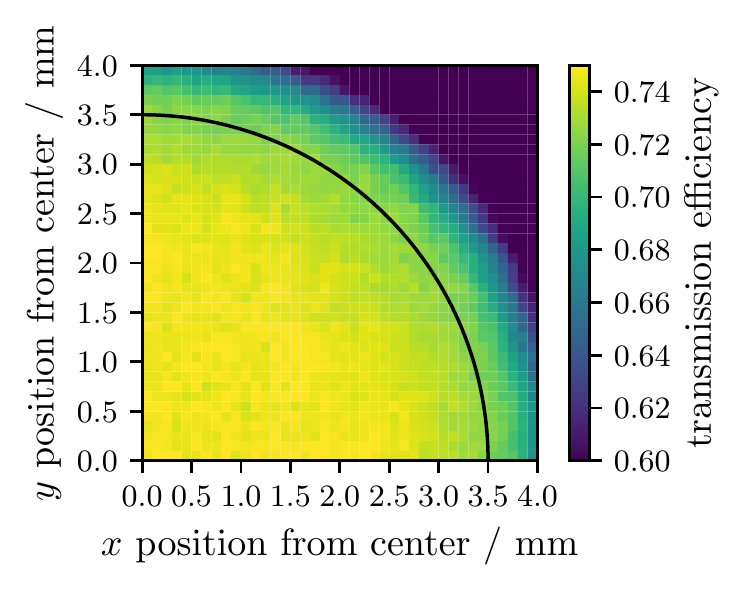}
			\caption{The Winston-shaped light guide.}
			\label{fig:FiberPosition_Winston}
		\end{subfigure}
		\caption{The transmission efficiency of the light guides for different fiber positions. The black line shows the maximal position the fibers can have inside an 8\,mm tube.}
		\label{fig:FiberPosition}
	\end{figure}

\begin{figure}\centering
    	\begin{minipage}[t]{0.49\textwidth}\centering
    		\includegraphics[]{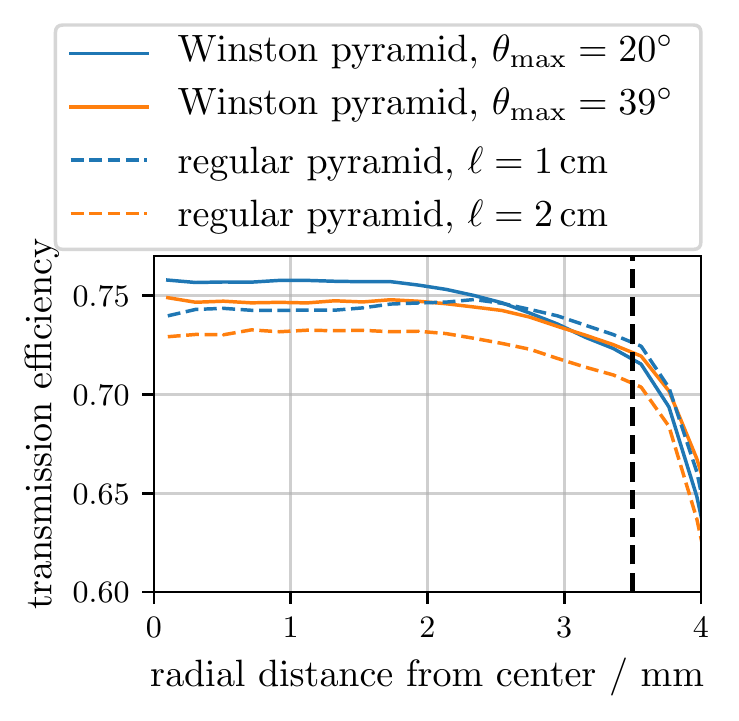}
			\caption{Mean efficiency depending on the radial distance of the fiber from the center for different light guide geometries. The dashed vertical line denotes the maximum radial distance a fiber can have inside the cookie.}
			\label{fig:FiberRadialDistance}
    	\end{minipage}
        \hfill
        \begin{minipage}[t]{0.49\textwidth}\centering
        	\includegraphics[]{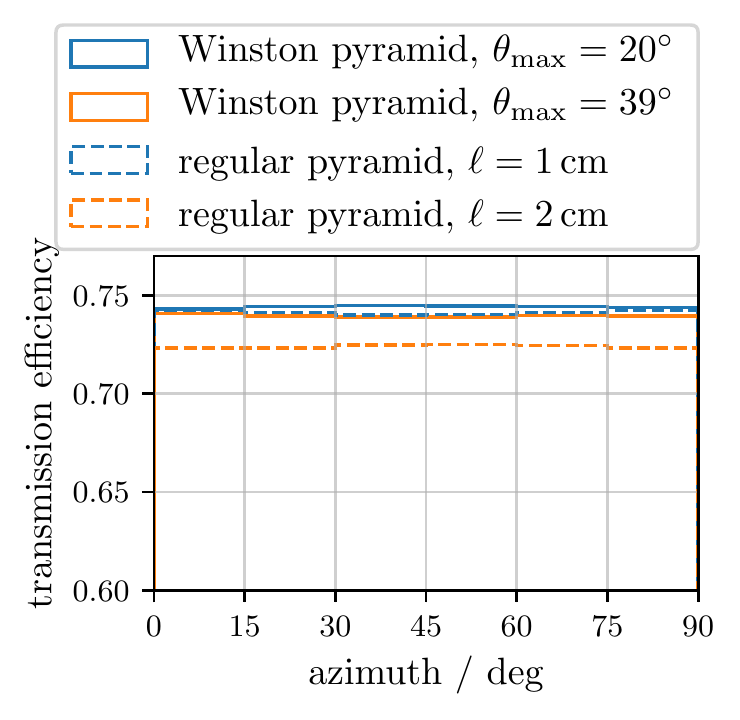}
            \caption{Mean efficiency depending on the azimuth angle $\varphi=\arctan(y/x)$. Only radial distances up to the maximum of \SI{3.5}{\mm} are taken into account. The distribution is flat for all tested geometries.}
            \label{fig:FiberAzimuth}
        \end{minipage}        
	\end{figure}

	To allow for a homogeneous detector response, the light transmission efficiency of the light guide should be independent of the fiber position in the fiber bundle. In figure~\ref{fig:FiberPosition}, scans of the fiber position are shown. This simulation is performed with only one fiber emitting light.
	The two light guides have a similar performance. In figure~\ref{fig:FiberRadialDistance}, the mean transmission efficiency is shown depending on the radial distance of one fiber from the center of the fiber bundle. The regular-shaped light guide with a length of 1\,cm has the lowest radial dependency at the order of \SI{+-1}{\percent} while for the other geometries it goes up to \SI{+-2}{\percent}.
    
    As the light guides are not rotationally symmetric, the azimuthal dependence of the transmission efficiency is tested in figure~\ref{fig:FiberAzimuth}. All geometries perform well and the dependency is below \SI{0.5}{\percent}.
    
    The regular-shaped light guide with a length of \SI{1}{\cm} performs best in terms of homogeneity for different fiber positions in the bundle.
    
    \subsubsection{Homogeneity of SiPM illumination}
    As SiPMs are cell structured devices, of which each cell can only detect a single photon in a time interval, inhomogeneities in the illumination alter its response. To achieve the highest dynamic range, each photon should have equal probability to hit any cell.
    \begin{figure}
    	\begin{subfigure}[t]{0.5\textwidth}\centering
    		\includegraphics[]{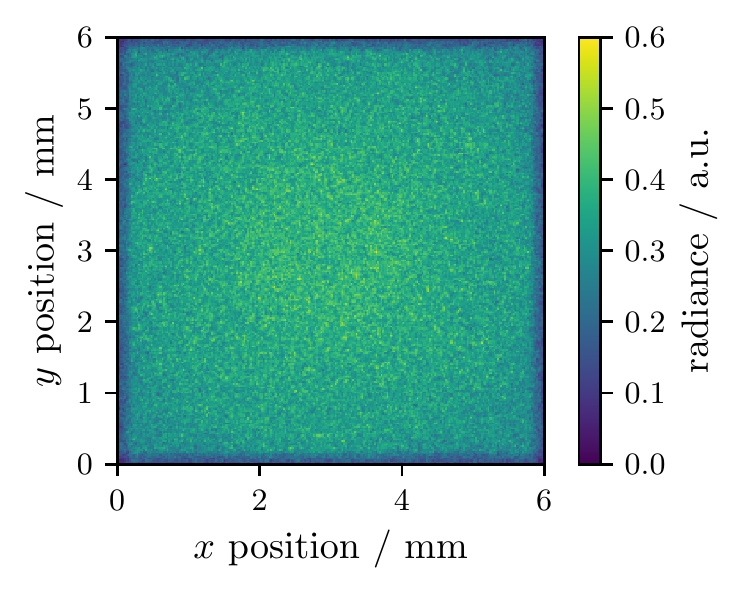}
            \caption{The regular-shaped light guide.}
    	\end{subfigure}
        \begin{subfigure}[t]{0.5\textwidth}\centering
        	\includegraphics[]{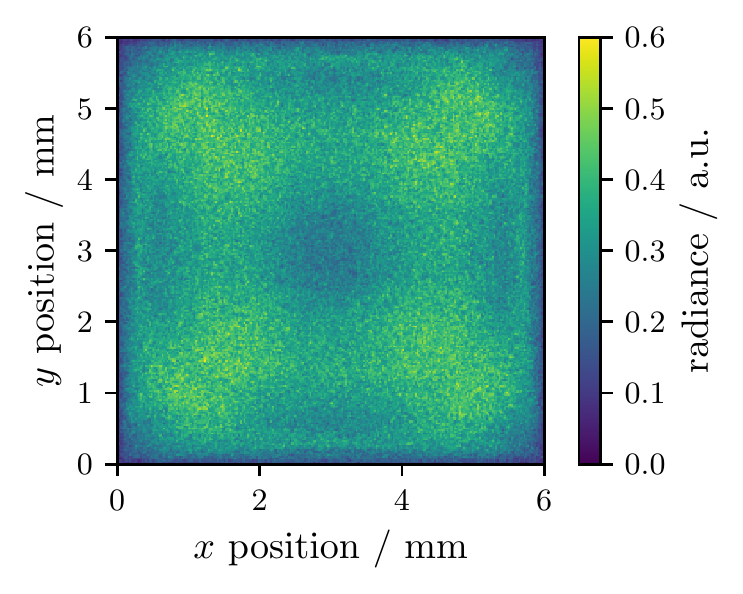}
            \caption{The Winston-shaped light guide.}
        \end{subfigure}
        \caption{The relative radiance of the light guides for each position on the SiPM. For this simulation, all fibers of the fiber bundle emit an equal amount of light.}
        \label{fig:ConeHomogeneity}
    \end{figure}
    In figure~\ref{fig:ConeHomogeneity}, a simulation of the light distribution on the SiPM is shown. The simulation is performed with all fibers of one bundle emitting the same amount of light. For both light guides, a narrow band of \SI{\sim0.15}{\mm} at the borders of the SiPM is not illuminated. To allow for some misplacement of the light guide, its size is retained.
    
    For the regular-shaped light guide, about \SI{13}{\percent} of the light is concentrated in the central \SI{2x2}{\mm} of the SiPM corresponding to about \SI{11}{\percent} of the area. This yields a slightly inhomogeneous distribution.
    
    The Winston-shaped light guide shows a higher illumination at the corners of the SiPM compared to its center. In the \SI{2x2}{\mm} area of all corners of the SiPM, \SI{45}{\percent} of the light is concentrated while it occupies \SI{44}{\percent} of the area.
    
    \begin{figure}\centering
    	\includegraphics[]{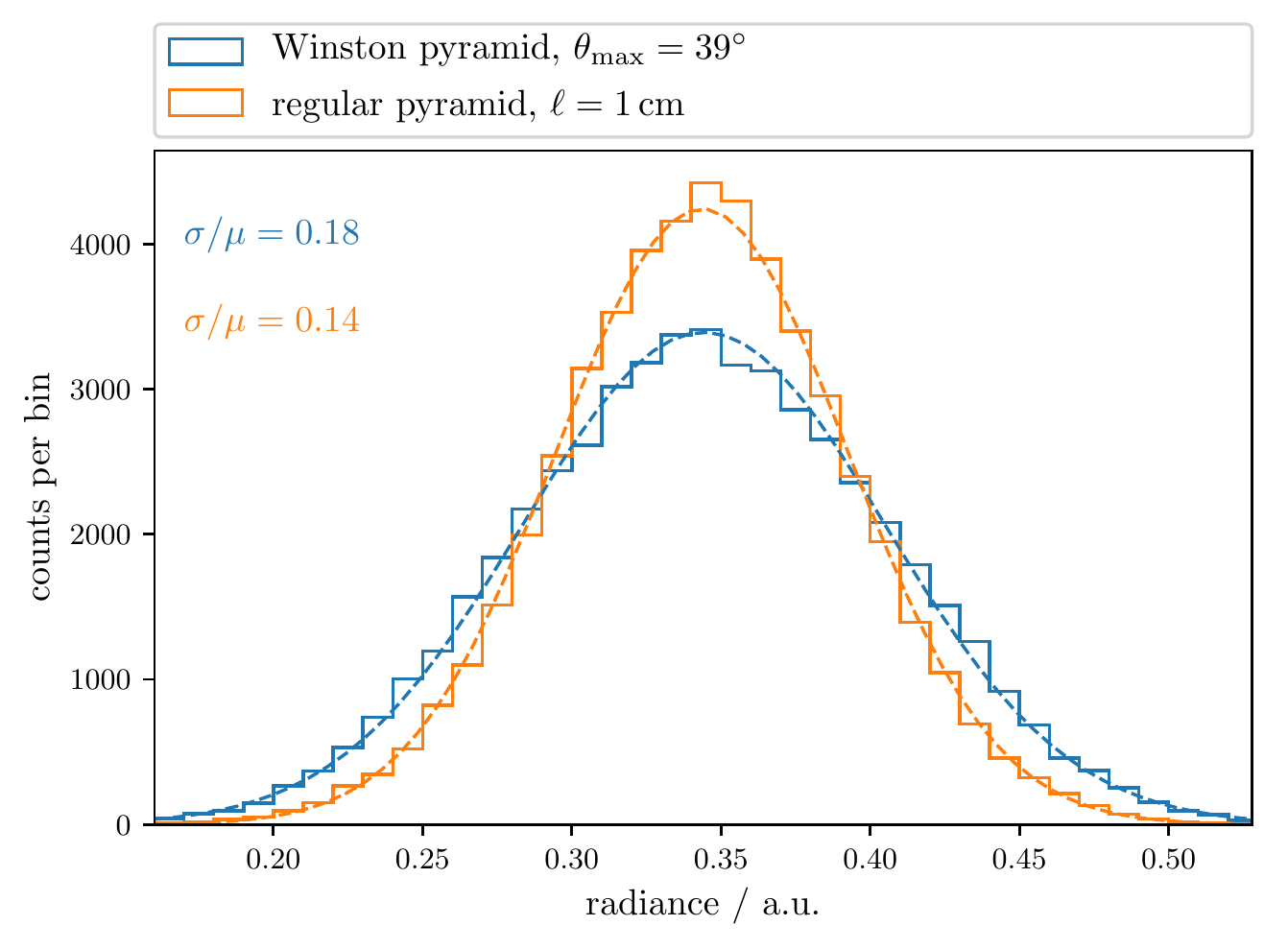}
		\caption{The spatial homogeneity of the illumination of the SiPM for two different light guide geometries. The simulation statistics is the same for both light guides. The simulated distributions show good agreement with Gaussian fits. $\sigma$ and $\mu$ refer to the Gaussian width and mean of the distribution, respectively.}
		\label{fig:ConeHomogeneity1D}
    \end{figure}
    In figure~\ref{fig:ConeHomogeneity1D}, the distribution of the radiance is shown. The simulated photon flux is the same for both geometries. Thus, purely statistical fluctuations are the same. The outer band with a width of \SI{0.1}{\mm} is not considered. For both distributions, a Gaussian fit agrees well with the data. The width of the distribution for the regular-shaped light guide is \SI{4}{\percent} narrower than that of the Winston-shaped light guide. 
    
    The most homogeneous distribution on the SiPM is produced by the regular-shaped light guide.
    
    \subsubsection{Conclusion}
    According to the simulation, all light guides show sufficient performance. The transmission efficiency for all light guides is above \SI{72}{\percent}. A light guide that consists of a truncated pyramid intersected with a cone with a length of \SI{1}{\cm} and an opening radius of \SI{4.7}{\mm} yields the best performance in terms of the most homogeneous efficiency for different fiber positions and also the most homogeneous light distribution on the light-sensitive surface of the SiPM.
    
    \subsection{Measurements}
    Different prototypes of the light guides have been produced by a dental technician~\cite{DentalLabor}. The prototypes were milled from PMMA and have hand-polished surfaces. Their transmission efficiency has been tested as well as the performance when installed in a Scintillator Surface Detector (SSD).
    
    \subsubsection{Light yield}
    The light yield was tested in a laboratory setup. The setup was designed to have optical properties as similar as possible compared to the SSD. Eight identical scintillator bars are read-out with optical fibers whose 16 fiber ends are guided onto a holding structure. The holding structure allows arranging the fibers in a way that their radial light distribution on the light guide is similar to the SSD setup with 48 fibers. Below the holding structure a light guide can be placed on a SiPM.
    
\begin{figure}\centering
    	\includegraphics[width=0.8\textwidth]{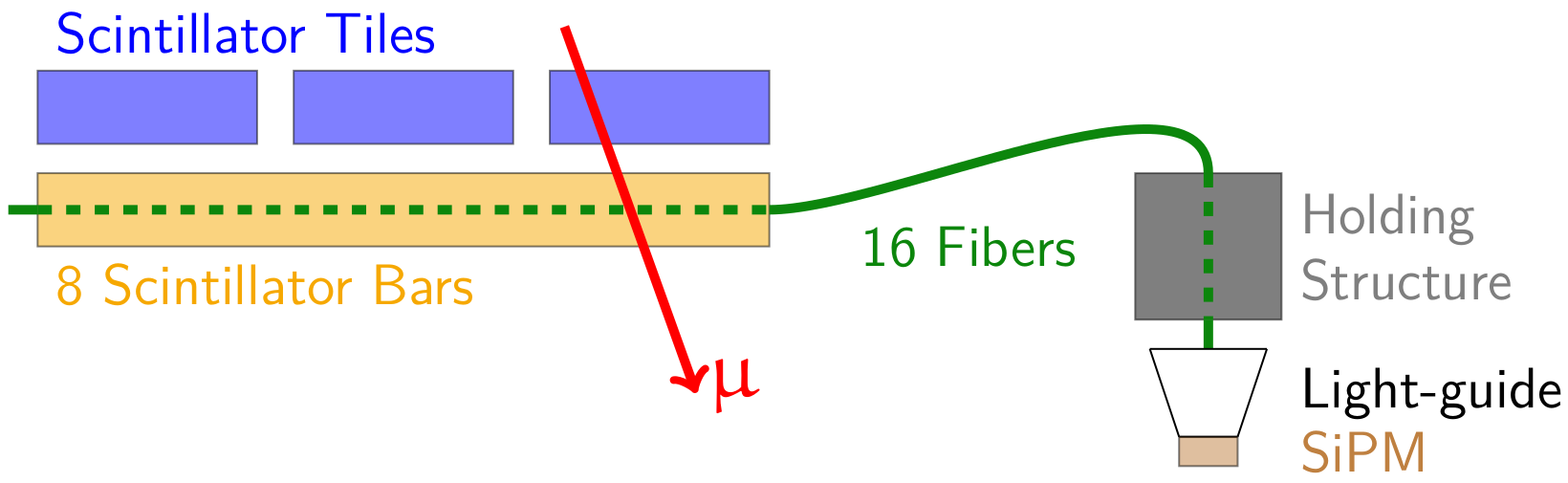}
        \caption{Schematic of the setup for determining the different performances of the light guides. The light produced in scintillator bars by through-going atmospheric muons is collected in WLS fibers and guided onto the light guide under test. It is read out with a SiPM and triggered by external scintillators.}
        \label{fig:ConeMeasurementSetup}
    \end{figure}

    Tests have been performed with light guides 
glued on the SiPM and with light guides only placed on the SiPM without special optical coupling material in between. The used glue (EPO-TEK 310M-2~\cite{EPOTEK}) has the same refractive index as PMMA and high optical transparency ensuring good optical coupling between the light guide and the SiPM.
    A schematic is shown in figure~\ref{fig:ConeMeasurementSetup}. 
    The light yield of minimum ionizing particles ("MIP signal") is measured for different light guides by triggering on atmospheric muons passing through the scintillators. A layer of scintillator tiles has been installed on top of the detector under test. The details of the tiles and their readout system can be found in~\cite{AMDICRC2015}. Coincidences between both detectors suppress the SiPM dark counts and noise originating from environmental radioactivity.
    
\begin{figure}\centering
    	\includegraphics[]{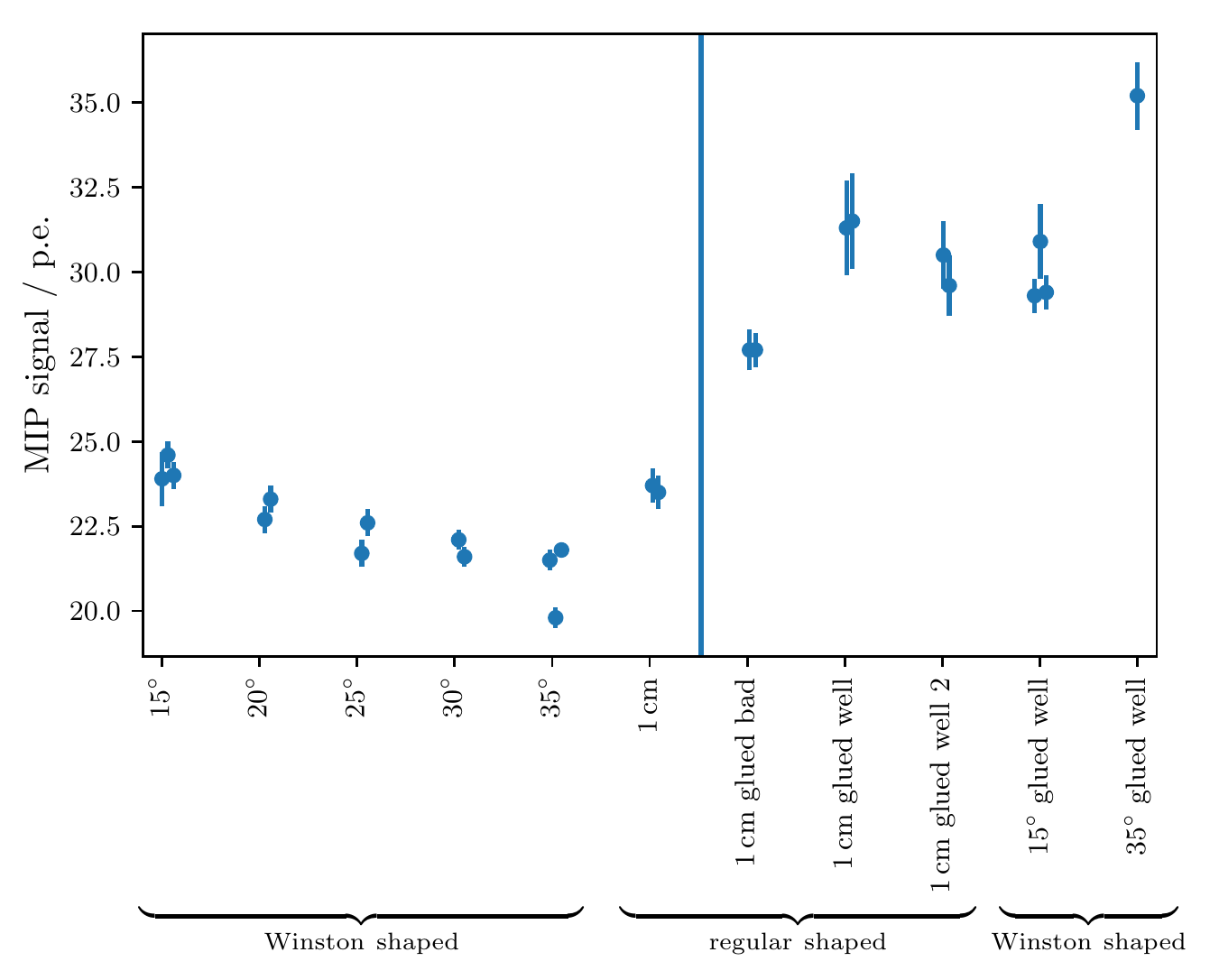}
        \caption{Measured light yield of different geometries of the light guides. The light yield is calibrated in photoelectron equivalents (p.e.). Light guides referred to by their length are regular-shaped while light guides referred to by cut-off angle are Winston shaped. At the right of the blue vertical line, measurements performed with light guides glued on the SiPM are shown. For the measurements on the left, the light guide is simply placed on the SiPM. Different data points for the same light guides represent repeated measurements after reassembling the setup to determine the systematic uncertainty.}
        \label{fig:ConeLightYieldMeasurement}
    \end{figure}

    In figure~\ref{fig:ConeLightYieldMeasurement}, the resulting light yield measured for different light guides is shown.
    A significant difference between the glued and none glued light guides is observed. A light guide that was badly glued on the SiPM shows a \SI{12}{\percent} worse performance than those that were glued well. The bad glue contains air-bubbles at roughly half the surface area. In total, only five light guides have been glued as it is an irreversible process and there were not enough SiPMs available to glue all produced geometries.
    
    For reflections at the unglued junctions between light guide and SiPM, the reflection coefficient increases with the incident angle and reaches total reflection for $\theta>74^\circ$. Without glue the light yield should thus increase with the average angle of the photons leaving the light guide. Long light guides (or small cut-off angles) result in smaller output angles and should perform better. The result as seen in figure~\ref{fig:ConeLightYieldMeasurement} shows exactly this behaviour for light guides that are not glued.
    
    As the glue has the same refractive index as PMMA of the light guides, the reflection coefficient and its angular dependency are lowered for glued light guides. For a cut-off angle of \ang{35}, the light yield increases by \SI{67}{\percent} when being glued on the SiPM, see fig.\ \ref{fig:ConeLightYieldMeasurement}. With a cut-off angle of only \ang{15}, the increase is only \SI{24}{\percent}. This result matches the expectation.
    
    All light guides that have been tested show sufficient performance for the application in a scintillator detector. The signal induced by a single through-going muon is well above the SiPM noise. In general, shorter light guides are favored due to lower absorption and lower probability for a photon to hit a surface of the light guide. Thus, surface imperfections have a lower impact on the performance of the light guide.
    
    \subsubsection{Spatial homogeneity of a full detector}
    To test the influence of the light guide on the homogeneity of a detector, the module has been installed in a prototype SSD designed for AugerPrime. The light guide tested here is the regular-shaped light guide with a length of 1\,cm. The detector was placed inside and triggered by a muon tomograph that tracks atmospheric muons with an accuracy of $\sim1\,$cm and has formerly been used as central tracker in the KASCADE experiment~\cite{KascadeMuonTracker}. Thus, the light yield for each position in the detector can be determined.
    \begin{figure}\centering
    	\includegraphics[]{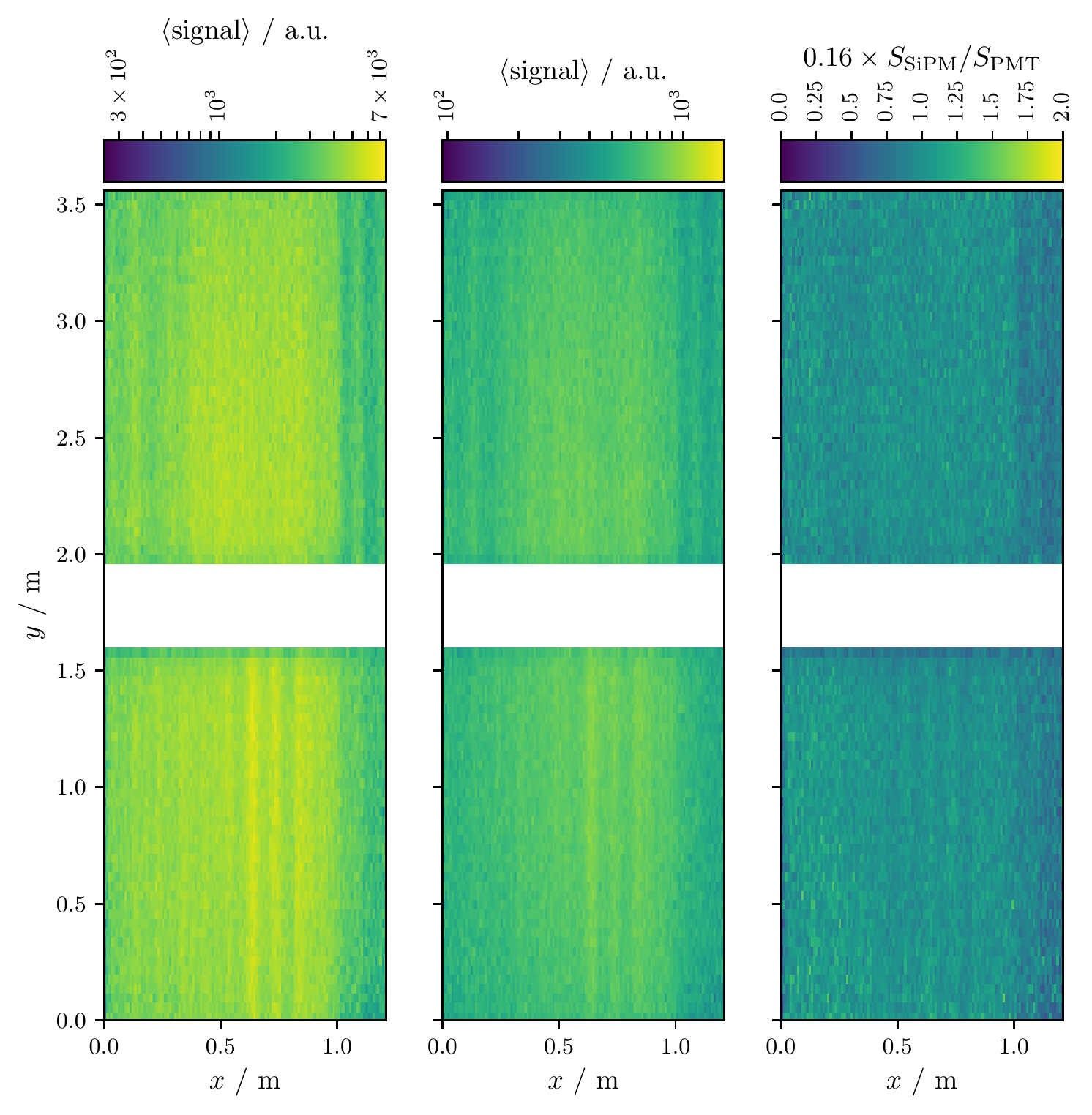}
        \caption{The light yield of the scintillator detector for muons crossing at different positions. The positional resolution for the crossing muons is \SI{1}{\cm}. As there is no strong y-dependence, a bin size of $1\times4$\,cm$^2$ has been chosen to reduce statistical fluctuations. \emph{Left:}~A measurement performed with the SiPM module. Some inhomogeneity is visible. \emph{Middle:}~A measurement performed with a PMT in the same detector. The inhomogeneity is similar as for the SiPM. Note the different color scales due to different SiPM and PMT signal sizes. \emph{Right:}~The ratio of the average signal measured for each position in the detector with the SiPM compared to the PMT. The SiPM signal was scaled such that the average ratio equals 1. The distribution is flat. Only the bars at the very right have some deviation resulting from production issues (c.f.\ fig.\ \ref{fig:SiPMFiberBundles}).}
        \label{fig:SSDDetectorHomogeneity}
    \end{figure}
    
    The measurement has been performed twice, once with the SiPM module and once with a conventional PMT installed in the same module. The result is shown in figure~\ref{fig:SSDDetectorHomogeneity}. The detector shows some spatial inhomogeneity. As it was a prototype, they result from imperfections in the manufacturing procedure and have been improved in later versions. Especially at the very right of the detector, the bars show a significantly lower signal.
    \begin{figure}\centering
    	\includegraphics[width=0.7\textwidth]{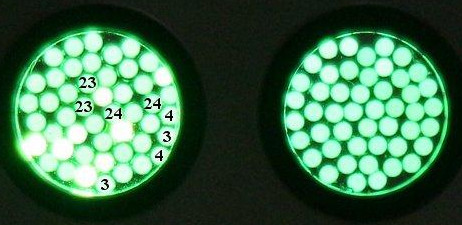}
        \caption{The two fiber bundles for the detector shown in figure~\ref{fig:SSDDetectorHomogeneity}. The left side corresponds to the top half ($y$=2.2 to 3.7\,m) of the detector. The fibers labelled 3 and 4 correspond to the bars at the very right of the detector. The fibers labelled 23 and 24 correspond to the bars at the left of the detector.}
        \label{fig:SiPMFiberBundles}
    \end{figure}
    
    As visible in figure~\ref{fig:SiPMFiberBundles}, the fibers routed through these bars are at the edge of the fiber bundle. In addition, they are shifted slightly back so that the distance to the light guide is larger. Thus, the worse light yield of the bars is explained by light that does not hit the light guide or PMT.
   
   On the right of figure~\ref{fig:SSDDetectorHomogeneity}, the ratio of the SiPM signal and the PMT signal is shown. The distribution is very flat, only the bars at the very right show a deviation. As described in figure~\ref{fig:SiPMFiberBundles}, the corresponding fibers are poorly positioned. The PMT has a relatively higher signal for these bars than the SiPM. Due to its large cathode size, it can still detect stray-light which would not be collected by the light guides. During production of future detectors, care will be taken to correctly position the fibers and to avoid this effect.
	
    \begin{figure}\centering
    	\includegraphics[]{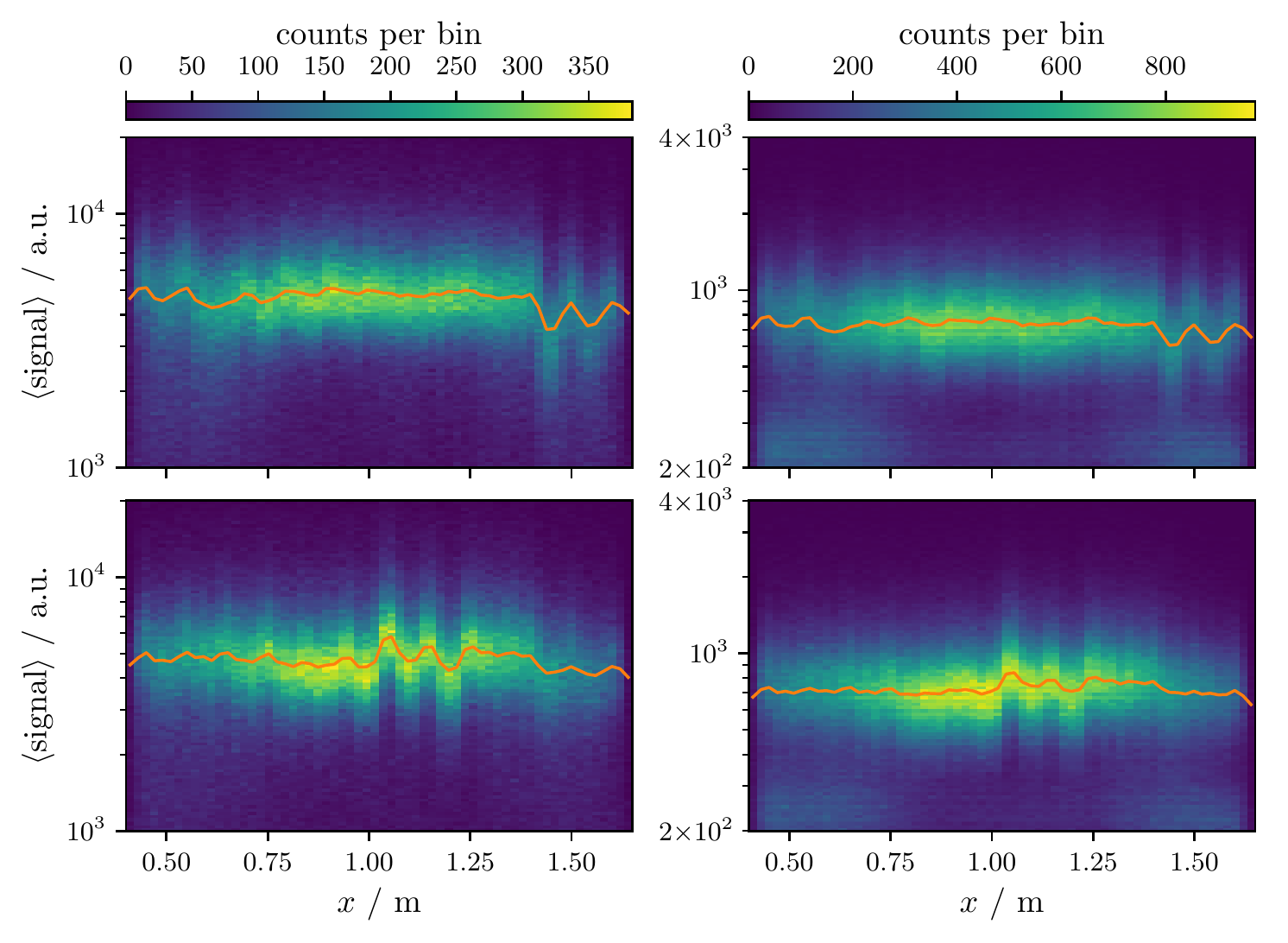}
        \caption{The signal distribution for different positions along the width of the detector as shown in figure~\ref{fig:SSDDetectorHomogeneity}. The \emph{left} plots correspond to the measurement with the SiPM module, the \emph{right} ones to that with the PMT. The \emph{top} and \emph{bottom} row show the signal for the two detector halves, respectively. Different scintillator bars can be identified because of the varying light yield. The median of the distribution is denoted by an orange line.}
        \label{fig:SSDBarInhomogeneity}
    \end{figure}
    In figure~\ref{fig:SSDBarInhomogeneity}, the light yield of different scintillator bars can be seen. In the top graphs, two bars  show a signal that is about \SI{30}{\percent} lower than the rest of the detector. In the bottom graphs, in the center of the detector, two bars have a significantly better light yield.
    \begin{figure}\centering
    	\includegraphics[]{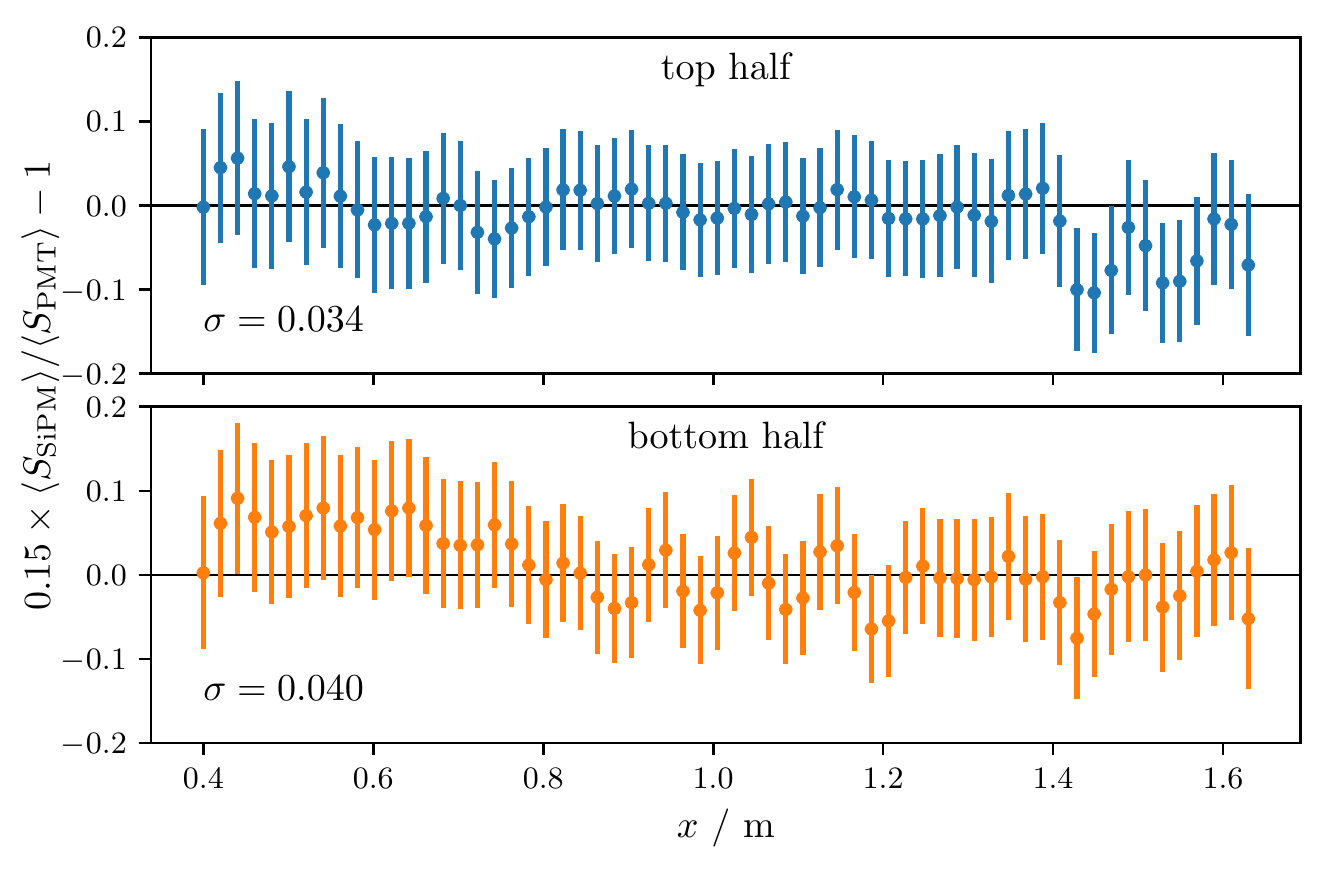}
        \caption{The ratio of the average signal for the SiPM and PMT normalized to equal means for the two detector halves. The RMS for the corresponding distribution is given in the plot.}
        \label{fig:SSDBarInhomogeneityRatio}
    \end{figure}
    
    The ratio of the average signal measured with the SiPM and with the PMT for each position is given in figure~\ref{fig:SSDBarInhomogeneityRatio}. The fluctuations are below \SI{10}{\percent} even for the poorly positioned fibers. The RMS of the corresponding distribution is $\SI{\leq4}{\percent}$. Thus, the light guides contribute to a detector inhomogeneity only at the level of a few percent.

\section{Electronics}\label{sec:Electronics}

The purpose of the electronics is to supply the two SiPMs with a temperature-adjusted bias voltage and to amplify the SiPM signal with the requirement to serve the full dynamic range of 1\,p.e. up to more than \num{100000}\,p.e. The typical SiPM bias voltage is less than \SI{60}{\volt} and a precision of better than \SI{30}{\milli\volt} is desirable (\SI{1}{\percent} of \SI{3}{\volt} nominal overvoltage). The temperature dependence of the SiPMs that needs to be compensated is around \SI{60}{\milli\volt\per\kelvin}. The module needs to be supplied and controlled via USB (\SI{5}{\volt}) and must not exceed a power consumption of \SI{500}{\milli\watt}. The electronics of the SiPM module fulfils these requirements.\\

The electronics module which has been shown in figure~\ref{fig:overview} consists of two printed circuit boards (PCB). Both PCBs hold connectors and surface mounted devices. The circular PCB hosts the Samtec MEC1 connector which is optimized for high-frequency applications. It serves as the interface between both PCBs. The circular PCB also holds an analog temperature sensor, type LM\,94021 by Texas Instruments, RC low pass filters for the bias voltage and the two SiPMs described previously.\\

The components for the SiPM bias- and low voltage power supply, and the pre-amplifiers are mounted on the main board. A mini USB type B connector is used to link the SiPM module with any standard USB port. Also, the SiPM bias- and low voltage power supplies are sourced from this connector. Three U.FL radio-frequency connectors by Hirose Electric provide the output of the three pre-amplifier channels. One USB type A to mini USB type B and three U.FL to SMA pigtails are used to connect the module to a data acquisition system, e.g.\ to an oscilloscope.\\

The module is protected by a coating based on acrylic resin which offers electric isolation. All components are selected for a temperature range of at least \SI{-20}{\celsius} to \SI{+50}{\celsius}. This range was selected in relation to typical maximum annual outside temperature variations.

\subsection{SiPM selection}

As the signal of both SiPMs is summed up in an analog circuit and both SiPMs are supplied with the same bias voltage, the SiPMs were pre-selected by the manufacturer with less than \SI{1}{\percent} variation in gain. In a sample of 16 modules, the maximum gain difference was found to be \SI{\leq0.3}{\percent}. Each module can be identified with a serial number. Two SiPMs are soldered on the circular PCB with a precision of \SI{50}{\um} using a professional assembling machine~\cite{Kuttig2016}.

\subsection{Low voltage power supply}

All three pre-amplifiers are supplied by two symmetric \SI{\pm3.3}{\volt} voltage rails. The \SI{+5}{\volt} voltage from the USB bus is inverted and regulated to \SI{-3.3}{\volt} using a low-noise, linear-regulated switched-capacitor voltage-inverter (charge pump). The USB voltage is regulated down to \SI{+3.3}{\volt} using a linear regulator.\\

Due to the high power supply rejection ratio of the linear regulators, noise from the inverting charge pump and the USB supply is suppressed by at least \SI{30}{\deci\bel}, over a bandwidth of \SI{1}{\mega\hertz} and over the specified temperature range from \SI{-50}{\celsius} to \SI{+150}{\celsius}~\cite{LTC3260}. This results in less than \SI{1}{\milli\volt} peak-to-peak ripple noise on either of the \SI{\pm 3.3}{V} rails. Both low voltage rails are filtered by two single stage RLC low pass filters, reducing the peak-to-peak ripple to less than \SI{100}{\micro\volt} in the bandwidth up to \SI{100}{\mega\hertz}.\\

Low-noise supply-voltages are essential when amplifying two \SI{36}{\milli\meter\squared} SiPMs with \SI{25}{\micro\meter} cell pitch as the signal amplitude scales with the cell size.

\subsection{SiPM bias voltage supply}
\label{sec:PowerSupply}

\begin{figure}\centering
	\includegraphics[width = .49\textwidth, keepaspectratio]{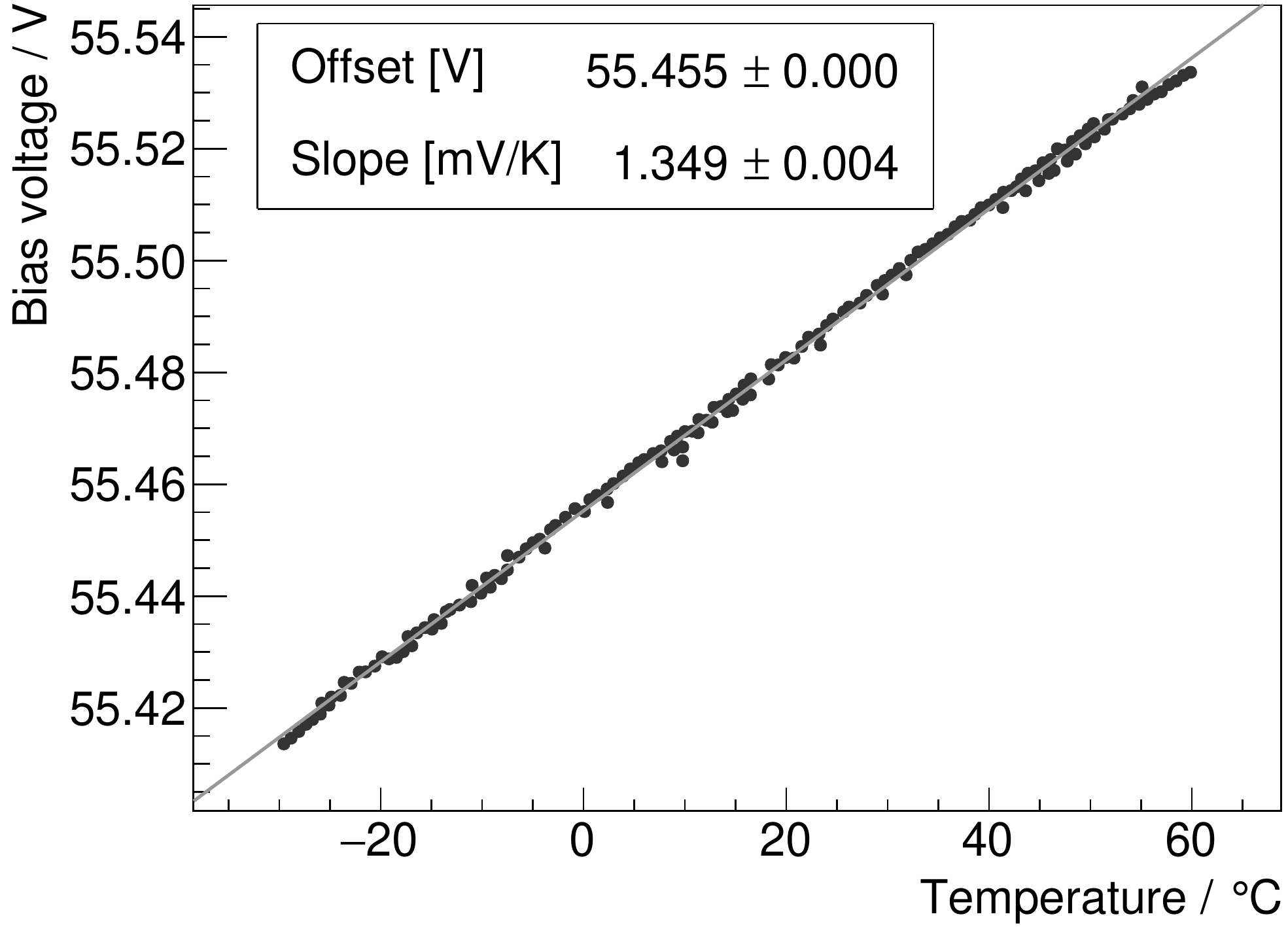}
	\includegraphics[width = .49\textwidth, keepaspectratio]{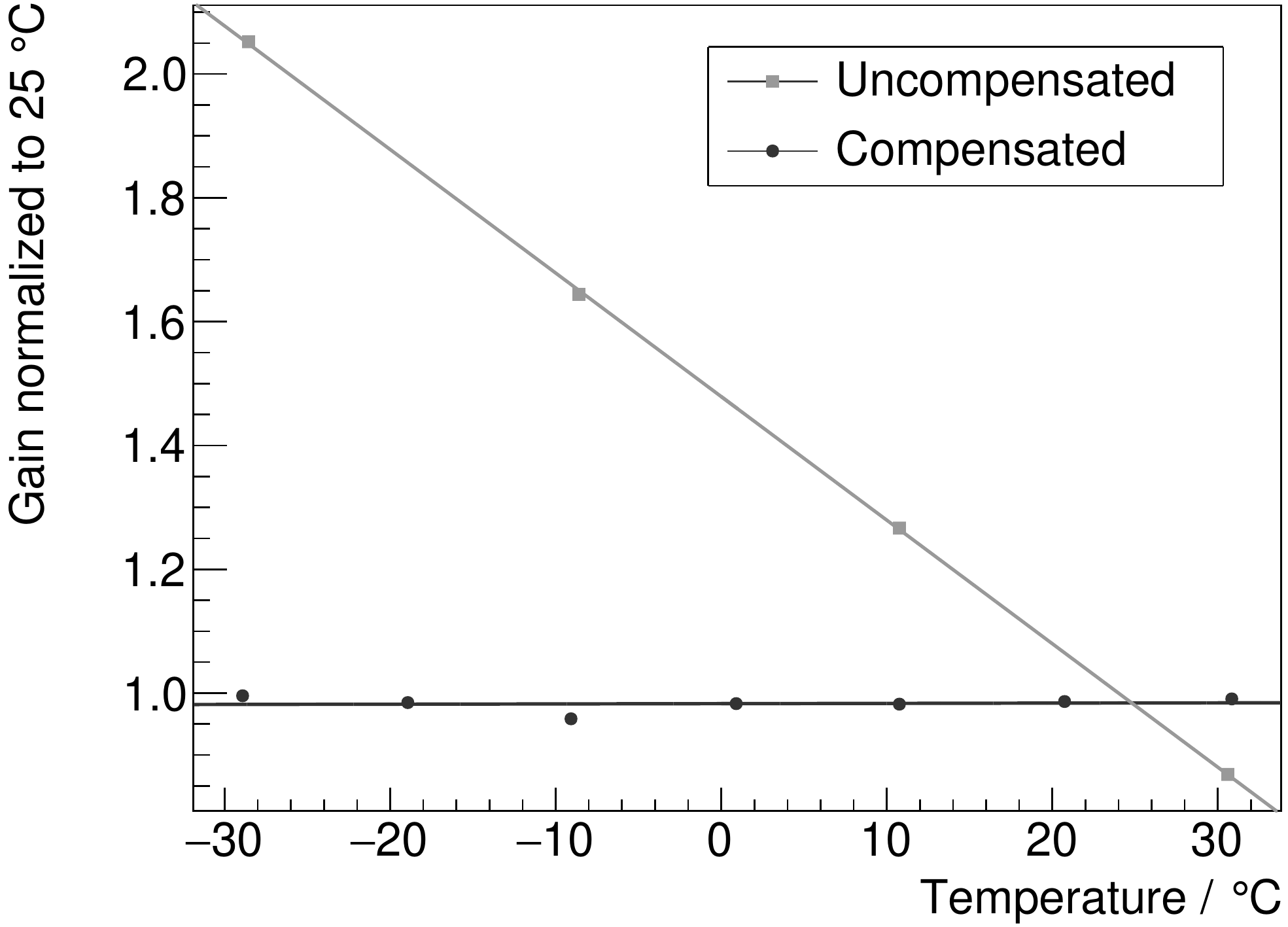}
\caption{\label{fig:temperature-dependence} \emph{Left:} Intrinsic temperature dependence of the Hamamatsu C11204-02 power supply chip as measured in a climate chamber. \emph{Right:} SiPM gain versus temperature compensated with a linear temperature progression and uncompensated with a constant bias voltage.}
\end{figure}

The SiPM bias voltage is generated and regulated by a commercial integrated circuit (IC), the Hamamatsu C11204-02, a digitally controlled power supply specialized for SiPMs~\cite{HamamatsuPSU}. The IC is powered via the \SI{5}{\volt} USB bus and features a DC/DC converter and regulator which generates the required bias voltage of about \SI{57}{\volt}. This voltage is programmable via the on-chip UART bus (Universal asynchronous receiver/transmitter, a simple serial interface) in \SI{1.8}{\milli\volt}-steps. The IC was tested in terms of temperature stability. A professional climate chamber was used for sweeping the ambient temperature around the IC from \SIrange{-30}{+50}{\celsius} while the output voltage was measured with a precision multimeter outside the climate chamber. For the two ICs, the intrinsic temperature dependence was found to be approximately \SI{1.3}{\milli\volt\per\kelvin}, see figure~\ref{fig:temperature-dependence} (left).\\

The IC allows to regulate its output voltage with a second-order polynomial with respect to the ambient temperature (measured by the analog temperature sensor) to correct for the temperature dependence of the breakdown voltage of SiPMs. The linear progression factor (about \SI{60}{\milli\volt\per\kelvin} for the used Hamamatsu SiPMs) and the second order parameter can be programmed into the IC via UART and the IC will apply the temperature-corrected bias voltage to the SiPMs automatically. The intrinsic temperature dependence of the IC must be taken into account. Figure~\ref{fig:temperature-dependence} (right) shows the SiPM gain determined from the SiPM charge histograms versus temperature. For one measurement, the SiPM was operated with a constant voltage and its gain is thus strongly temperature dependent. The gain, normalized to \SI{25}{\celsius}, changed by approximately $-\SI{2}{\percent\per\kelvin}$. To adjust the voltage according to the temperature, the Hamamatsu bias supply IC was programmed with a linear progression of approximately \SI{58}{\milli\volt\per\kelvin} which resulted in a remaining gain-temperature dependence of less than \SI{e-4}{\per\kelvin}, and thus less than \SI{1}{\percent} on the full \SIrange{-20}{+50}{\celsius} temperature range.\\

The applied voltage is accessible via UART with a digital resolution of \SI{1.8}{\milli\volt}, the current drawn by the SiPMs with a digital resolution of \SI{5}{\micro\ampere} and the temperature with a digital resolution of \SI{35}{\milli\kelvin}. The list of commands and their implementation can be found in~\cite{HamamatsuPSU}.

\subsection{USB control}

To communicate with the Hamamatsu C11204-02 digitally, the USB-to-UART bridge FT232R by FTDI is used \cite{FT232R}. The FT232R can be controlled with a symbol rate of \num{38400} baud, 8-bit, even parity and a single stop bit to meet the requirements of the C11204-02 UART communication interface. It was selected to operate its input and output lines with a \SI{5}{\volt} TTL level. To reduce electromagnetic interference during communication, the communication lines are physically isolated from the ground and power planes. Series resistors in the communication lines match the PCB trace impedance and reduce ringing and ground bouncing effectively.

\subsection{Pre-amplifiers}\label{sec:PreAmplifiers}

\begin{figure}
	\begin{center}
	\includegraphics[height = .35\textwidth, keepaspectratio]{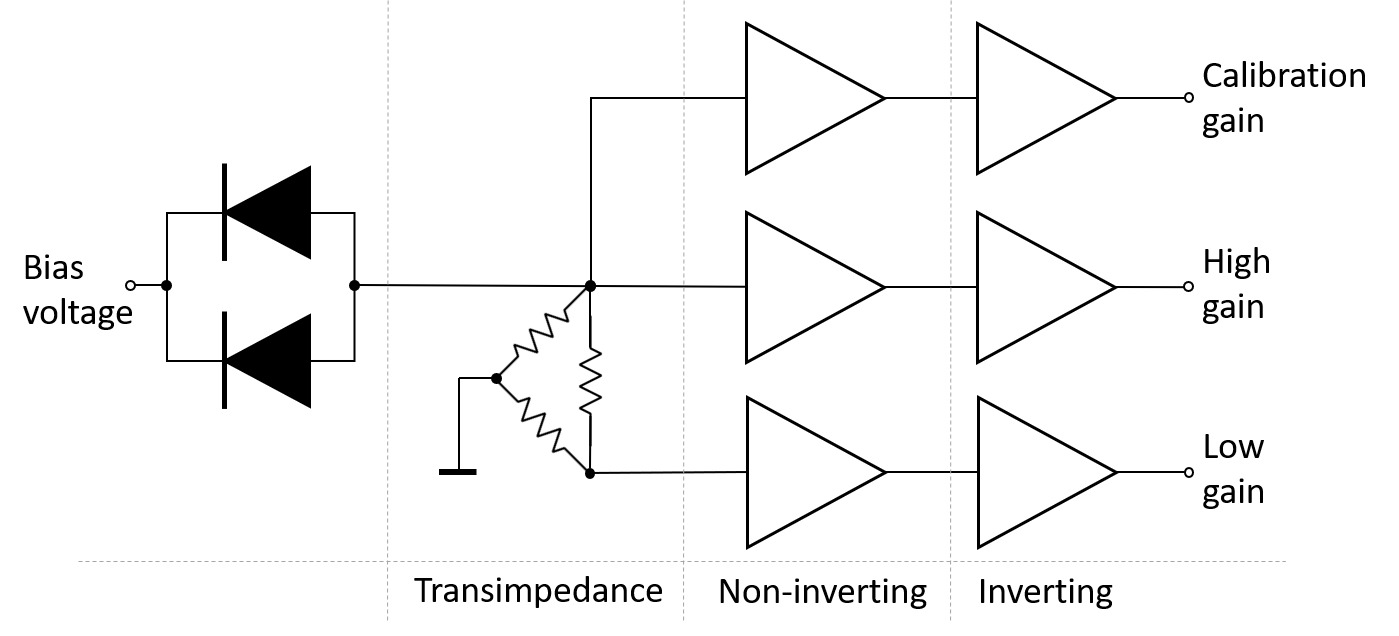}
	\end{center}
\caption{\label{fig:preamps} This figure shows the schematic overview of the pre-amplifier architecture used in the latest revision of the printed circuit board. Two SiPMs are connected in parallel and read-out by a simple resistor bridge (transimpedance stage) with a total impedance of \SI{50}{\ohm}. A non-inverting current feedback stage is followed by an inverting stage forming three individual amplifier channels.}
\end{figure}

Because of the limited resolution and noise floor of typical read-out systems, it has been decided to implement three individual gain channels on the front-end electronics to increase the signal-to-noise ratio. This improves calibration features of the SiPMs, especially the single p.e.\ resolution which is a characteristic quality of SiPMs.\\

\paragraph{Design} The pre-amplifier consists of a single transimpedance stage followed by three voltage amplification channels connected in parallel. The architecture of the latter is composed of a non-inverting and an inverting stage each, see also figure~\ref{fig:preamps}.\\

In the latest revision, the transimpedance stage was implemented with a simple bridge resistor circuit. The calibration channel and the high-gain channel are connected to the same potential with a \SI{50}{\ohm} transimpedance. The low-gain channel is connected to the divider node with a \SI{25}{\ohm} transimpedance. The (technical) current flows from the positive bias voltage junction of the SiPMs through the resistor bridge to ground. This introduces a voltage drop on the transimpedance nodes which can be amplified by standard amplifier circuits. The desired (application-specific) output polarity of the three gain channels is chosen to be negative. So, each two-stage amplifier channel needs to host an inverting and a non-inverting amplifier. A current-feedback amplifier (CFA) integrated-circuit, type Analog Devices AD8012, was selected with a maximal small-signal bandwidth of \SI{350}{\mega\hertz}, a slew rate of \SI{2250}{\volt\per\micro\second}, a low power consumption of less than \SI{1.8}{\mA} per amplifier and a high drive level of nominal \SI{100}{\milli\ampere}. With CFAs, the non-inverting stage needed to be placed before the inverting stage because of their low inverting input impedance and the high feedback currents for large pulses. Placing the inverting stage before the non-inverting stage would introduce distortion between the channels, especially when the calibration channel and/or high-gain channels got saturated. The non-inverting amplifier architecture profits from a very high input impedance and a feedback path which is de-coupled from the input. With this implementation, the SiPMs only see the load impedance of the resistor bridge. This isolates the individual gain channels, reducing distortion significantly. The absolute transimpedance gain of the calibration, high and low-gain channels are approximately \SI{17200}{\ohm}, \SI{800}{\ohm} and \SI{25}{\ohm}, respectively.\\

\begin{figure}
	\includegraphics[width = .49\textwidth, keepaspectratio]{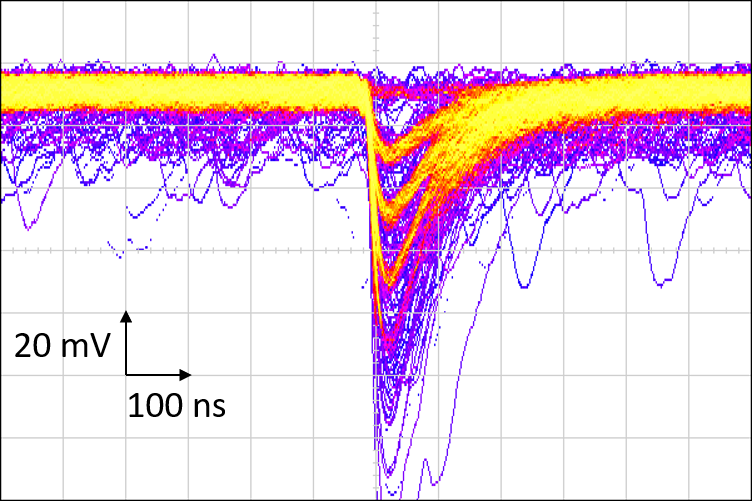}
    \hfill 
	\includegraphics[width = .49\textwidth, keepaspectratio]{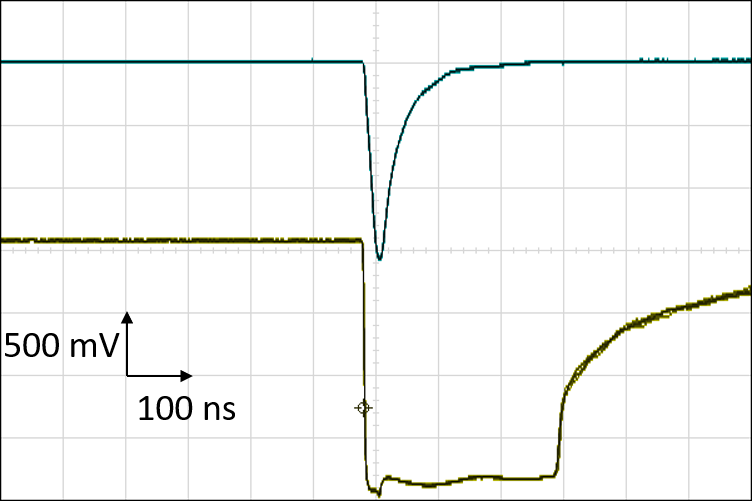}
\caption{\label{fig:preamps_signals} \emph{Left:} Two \SI{6x6}{\square\mm}-sized, \SI{25}{\um} cell pitch SiPMs amplified by the calibration channel flashed by and triggered on an LED with \SI{1}{\second} persistence. \emph{Right:} Large-signal response (>\num{90000}\,p.e.) of the high-gain and the low-gain channel using a pulsed LED. No undershoot or oscillation can be observed even though the high-gain and inevitably also the calibration channel are clearly saturated.}
\end{figure}

This design was tested in a dark chamber and with two \SI{6 x 6}{\milli\meter\squared} SiPMs. The bias voltage was set according to the specifications of the SiPMs. The left-hand side of figure~\ref{fig:preamps_signals} shows the response of the calibration channel where the SiPMs have been illuminated by a small light pulse. The read-out was synchronized with the trigger and a persistence of \SI{1}{\second} was set. The first photon equivalents are clearly visible in the calibration channel. To test the system at the upper limit of the dynamic range (see fig.\ \ref{fig:preamps_signals}, right), the module was illuminated by an LED that was pulsed by a large, and short (\SI{10}{\nano\second} FWHM) electric pulse. Even though the high-gain channel was saturated, no signal undershoot, or excessive ringing could be observed on the low-gain channel.

\subsection{Power consumption}

The current drawn during standard operation in the dark is about \SI{38}{\milli\ampere} on the \SI{+5}{\volt} USB supply rail thus dissipating \SI{190}{\milli\watt}. The test conditions were an operation of both SiPMs with a bias voltage of \SI{57.35}{\volt} and a load of less than \SI{1}{\uA} at \SI{22}{\celsius}.

Adding or removing single SiPMs will not measurably increase nor decrease the total power consumption in this configuration. On the other hand, changing the number of pre-amplifier channels will greatly influence the total power consumption with approximately \SI{30}{\milli\watt} per channel.

\section{Preliminary results from the field}

\begin{figure}\centering
	\begin{minipage}[t]{0.49\textwidth}\centering
		\includegraphics[width=\textwidth]{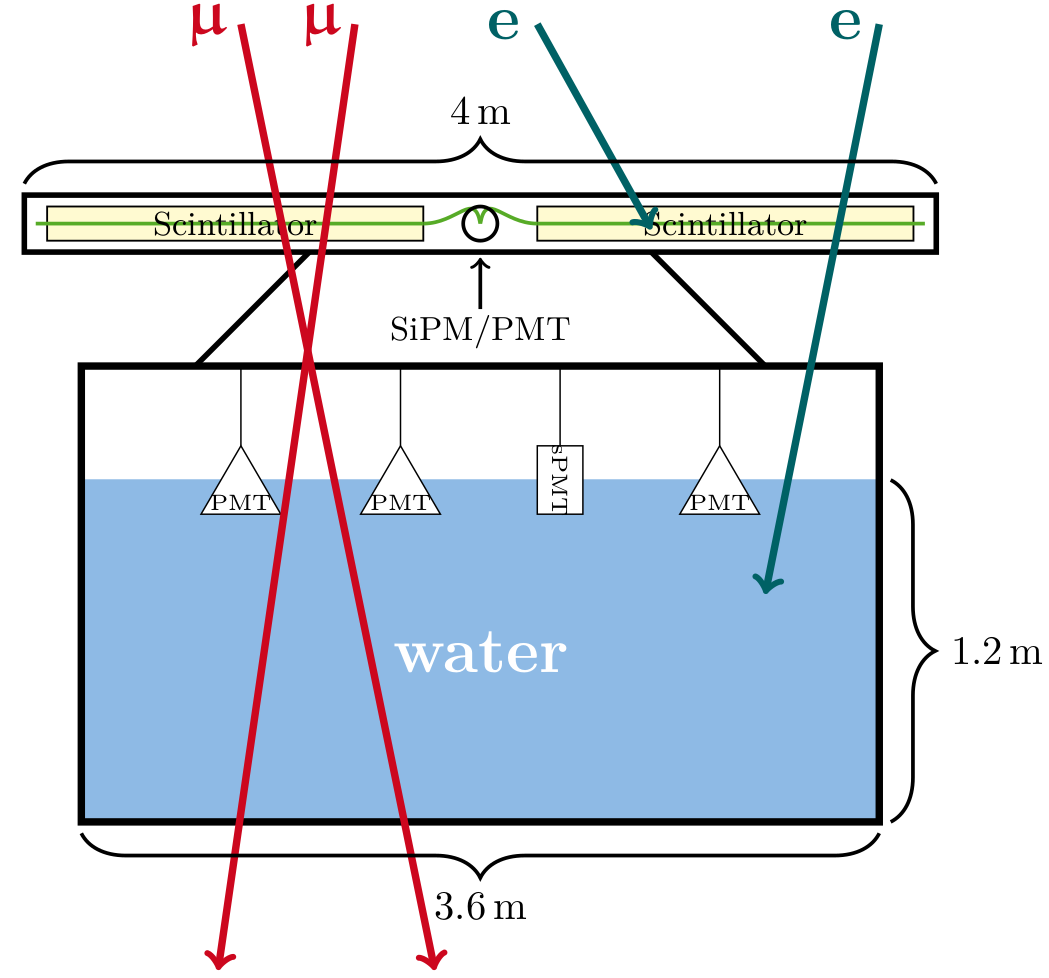}
    	\caption{A sketch of a Pierre Auger Detector Station after deployment of the SSD. The SSD is mounted on top of the water-Cherenkov detector and detects both the electromagnetic and the muonic shower component. According to the baseline design, it is read-out with a PMT. For test purposes, three detectors are equipped with a SiPM module.}
    	\label{fig:SSDonSDSketch}
	\end{minipage}
    \hfill
    \begin{minipage}[t]{0.49\textwidth}\centering
    	\includegraphics[width=\textwidth]{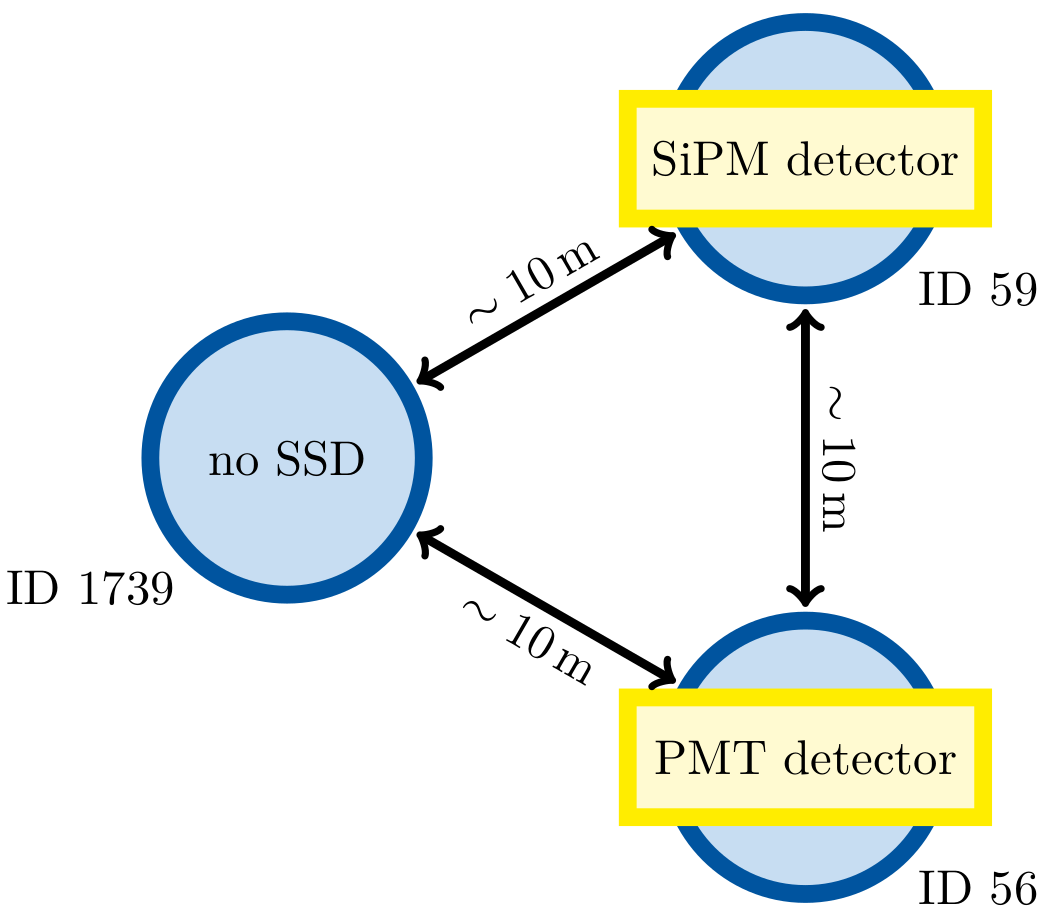}
        \caption{A schematic view of the station layout at the detector station equipped with a SiPM module. One neighbouring station is equipped with an SSD with a PMT and a third station is not equipped with an SSD. A unique ID identifies the stations.}
        \label{fig:EALayout}
    \end{minipage}
\end{figure}
\paragraph{Tests} The described optical module was tested in three prototype Scintillator Surface Detectors (SSD) which were deployed together with nine standard PMT based prototypes at the Pierre Auger Observatory in September 2016 on top of existing surface detector (SD) stations~\cite{EADeploymentMartello}. A schematic of such a station is shown in figure~\ref{fig:SSDonSDSketch}. 

Data from the operation allows understanding the performance of the SiPM module in a realistic application scenario. While usually stations are separated by \SI{1.5}{\km}, one of the SiPM stations has two neighbouring stations that are only about \SI {10}{\meter} apart. On top of one of these stations, a SSD equipped with a PMT is installed. The other station is a standard Auger detector station without a SSD prototype. A schematic of the layout is shown in figure~\ref{fig:EALayout}. 

The installed SiPM modules are equipped with a preliminary version of the pre-amplifiers described in section~\ref{sec:PreAmplifiers}. The power supply is the same as described in section~\ref{sec:PowerSupply}. Thus, the long-term stability of the module will be the same for the versions in the field and described here.

\subsection{Calibration histograms}
During standard operation of the detector, calibration histograms are taken allowing the determination of the charge measured for a through-going minimum ionizing particle (MIP). They are used to calibrate the light yield of the detector and to allow a comparison of different detector stations. The calibration histograms are only taken at one channel of the optical module.

During the first period of data taking, histograms are taken from the calibration-gain channel. This allows determining the gain stability of the SiPMs.
\begin{figure}\centering
	\includegraphics[]{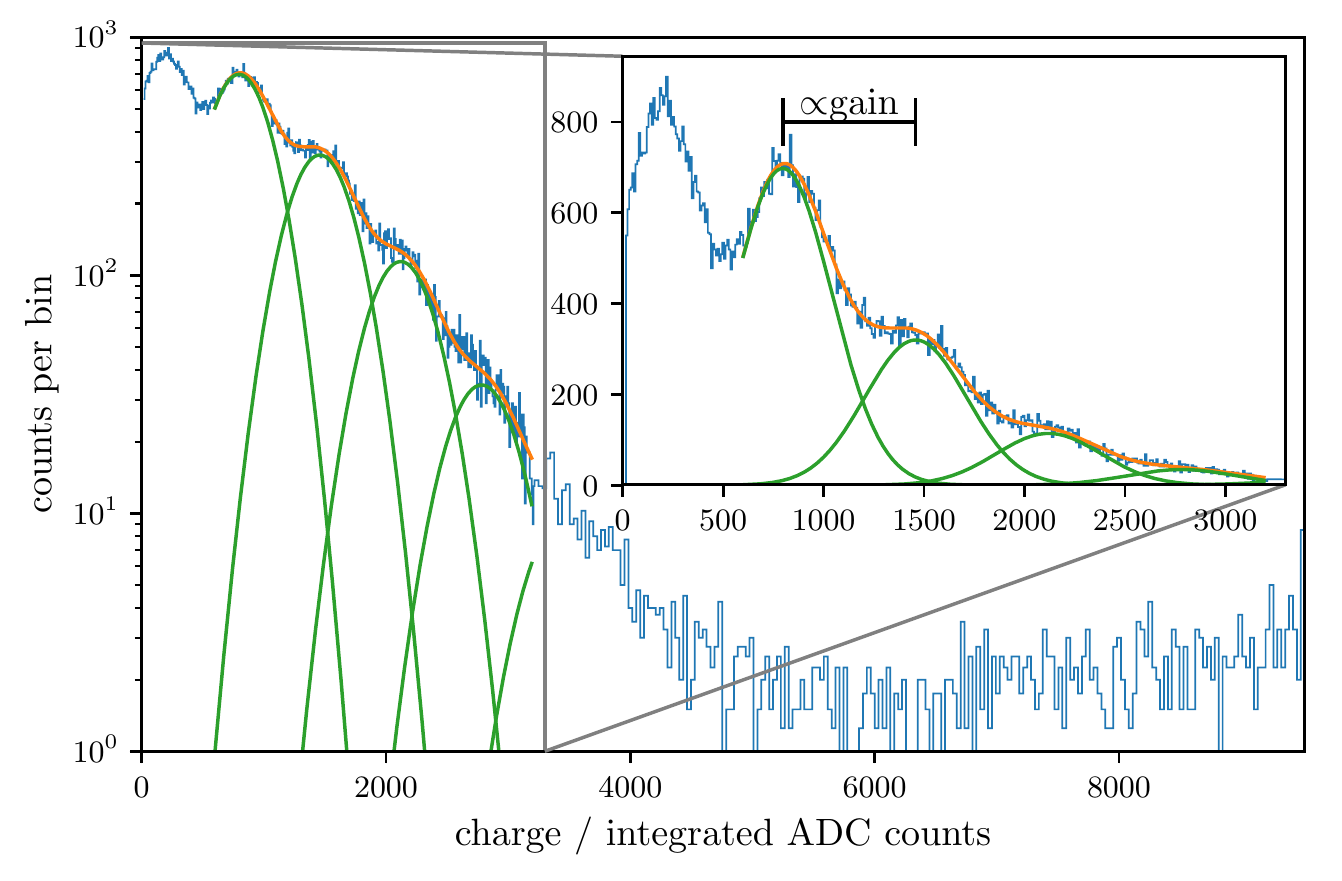}
    \caption{An example of a calibration histogram measured with the calibration-gain channel of the SiPM module. Due to the limited range of the histogram, the MIP peak at >30\,p.e.\ is not visible here. Peaks corresponding to a single p.e.\ are well visible and allow determination of the SiPM gain.}
    \label{fig:CalibrationHistoCG}
\end{figure}
In figure~\ref{fig:CalibrationHistoCG}, a calibration histogram is shown. Due to the limited range of the histogram, the MIP peak is not visible here.

Multiple correlated Gaussians were fitted to the first four peaks of the dark count spectrum. The distance between the Gaussians is proportional to the SiPM gain and thus allows monitoring it.
\begin{figure}\centering
	\includegraphics[]{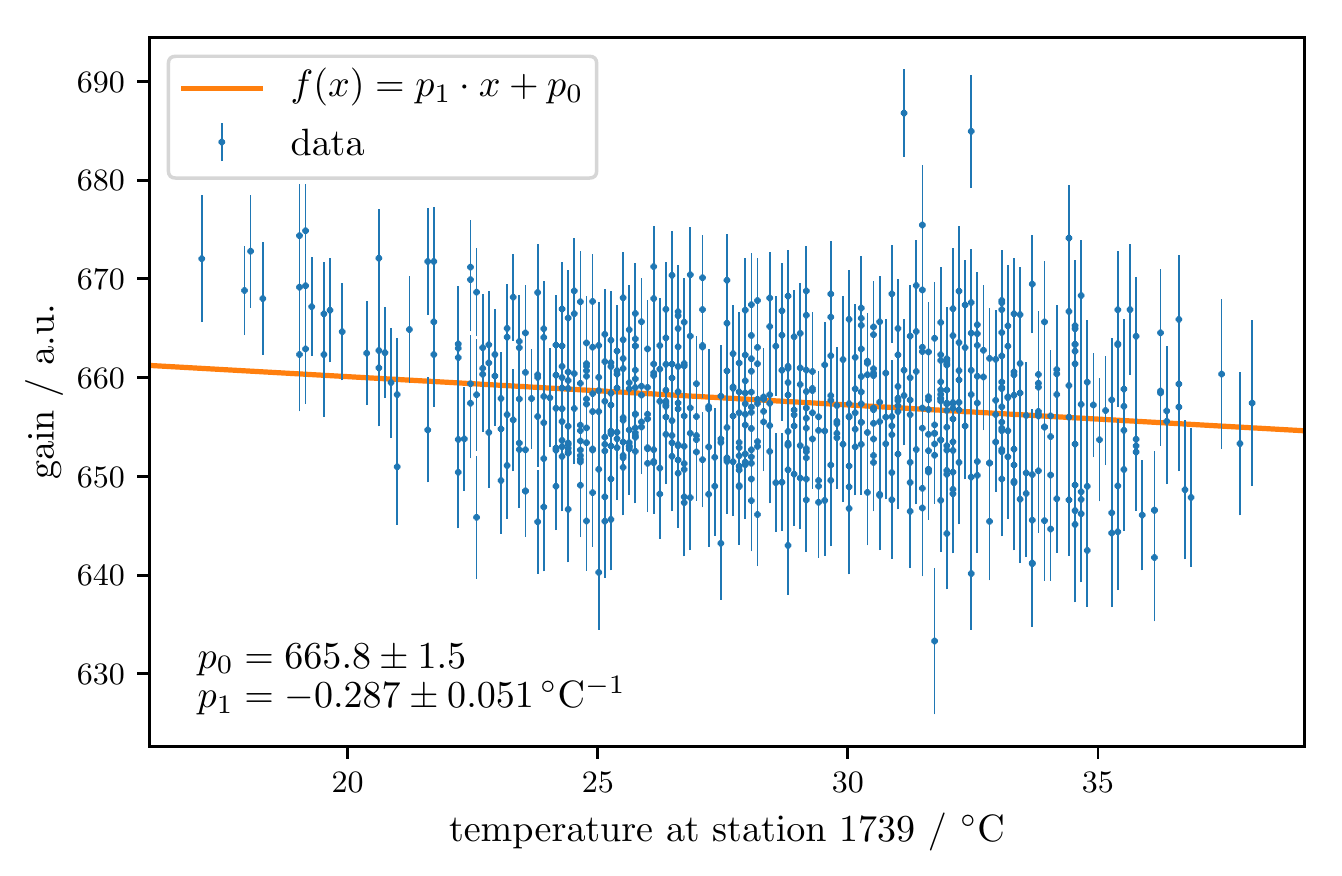}
    \caption{The gain of the SiPM with respect to the temperature. At the time of data taking, the read-out of monitoring values such as temperature was not yet implemented for the new detectors. Thus, the temperature was taken from the old station 1739 which is \SI{\sim10}{\meter} apart.}
    \label{fig:GainvsTemperature}
\end{figure}
The correlation between gain and temperature can be seen in figure~\ref{fig:GainvsTemperature}. When the data was taken, the read-out of monitoring values such as temperature was not yet implemented for the new detectors. Thus, the temperature measured at the neighbouring old station 1739 was taken instead.

The gain has been found to be as stable as \SI{0.04}{\percent\per\kelvin}. The module allows for a more precise setting of the temperature-dependent correction of the SiPM bias voltage. Thus, having measured the dependency, the correction will be adjusted.

\begin{figure}\centering
	\includegraphics[]{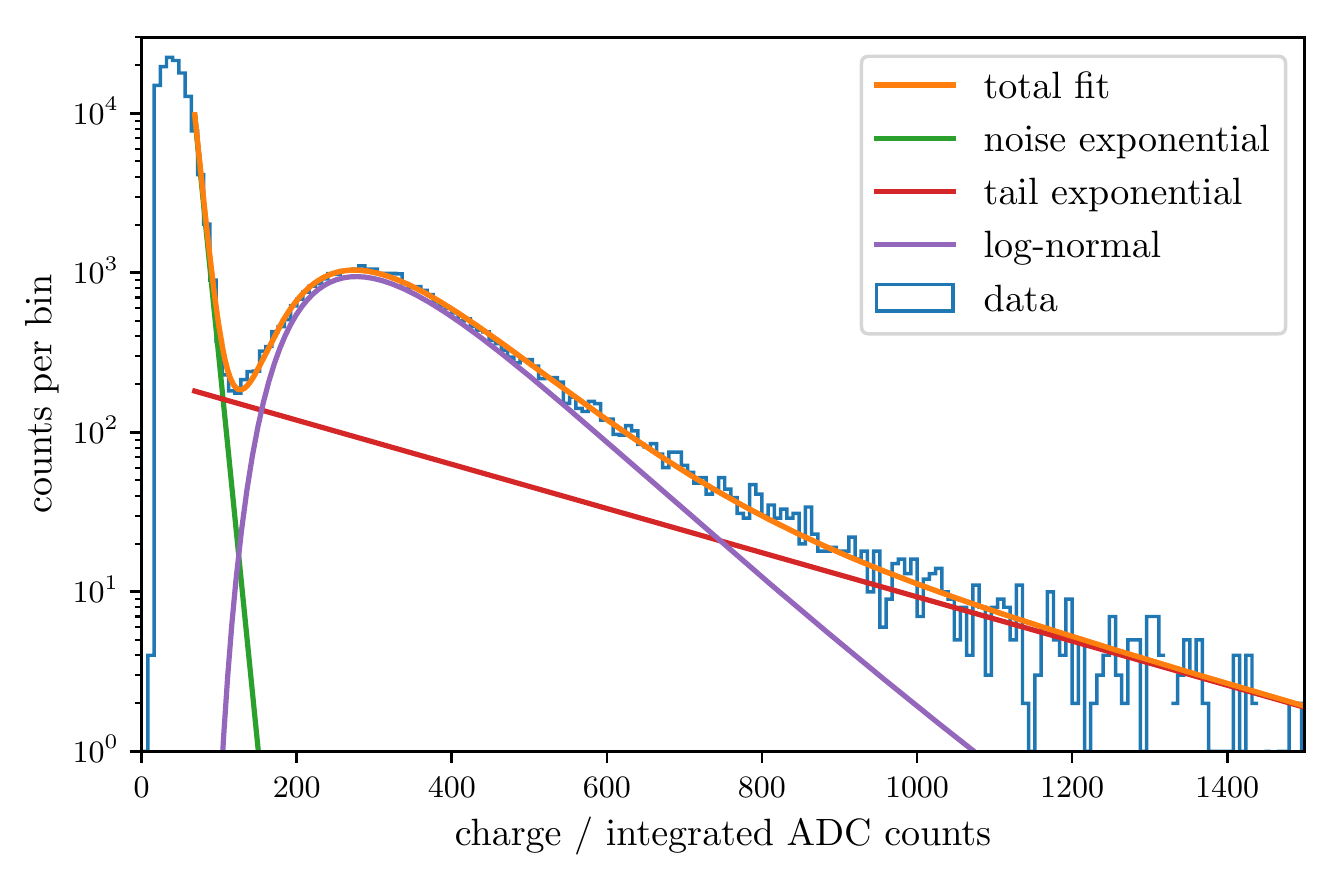}
    \caption{An example of a calibration histogram measured with the high-gain channel of the SiPM module. Here, the MIP peak is visible and well separated from the noise. The shown fit consists of three components. First, an exponential function describing the exponential SiPM noise background; second, an exponential to describe the tail of the distribution; third, a log-normal distribution to fit the peak.}
    \label{fig:CalibrationHistoHG}
\end{figure}

During another measurement period, the calibration histograms were taken at the high-gain channel. Such a histogram is shown in figure~\ref{fig:CalibrationHistoHG}. The spectrum is dominated by the light distribution produced by single minimum ionizing particles. The clearly visible peak is well fit by a log-normal distribution with an exponential tail. In contrast, a Landau distribution convolved with a Gaussian will show some discrepancy with the measurement. The fit allows to determine the most probable value of the spectrum and relative changes of the light yield can be derived. The peak is well separated from the noise and background with a peak-to-valley ratio of 5.6.

\begin{figure}\centering
	\includegraphics[]{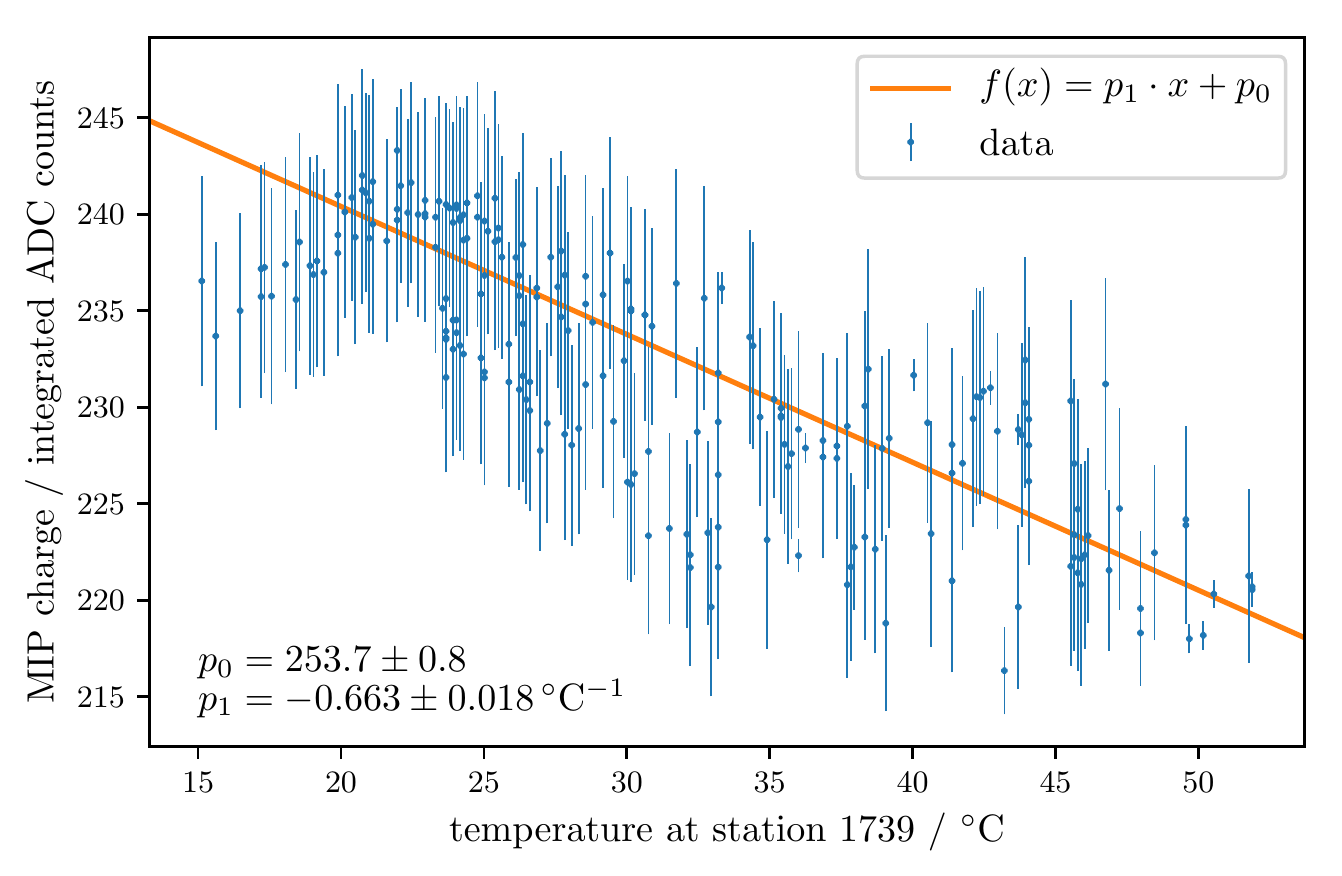}
    \caption{The correlation between the measured charge for a MIP and the temperature. As in fig.~\ref{fig:GainvsTemperature}, the temperature has been measured at a neighbouring station.}
    \label{fig:MIPvsTemperature}
\end{figure}

As can be seen in figure~\ref{fig:MIPvsTemperature}, there is a strong correlation between the measured charge for a MIP and the temperature. The dependency is \SI{0.28}{\percent\per\kelvin} which is clearly dominating over the small gain variation. As the gain of the SiPMs is stable, the temperature dependence of the MIP signal must originate from a change of the light flux. This effect most likely results from a temperature dependent light yield of the scintillators and fibers.

As the gain of the SiPM can be monitored in the field, the effects of a changing gain and a changing light yield can be disentangled. With a conventional PMT this is not possible during normal data taking operation.

\subsection{Events measured with the SiPM module}
\begin{figure}\centering
	\includegraphics[]{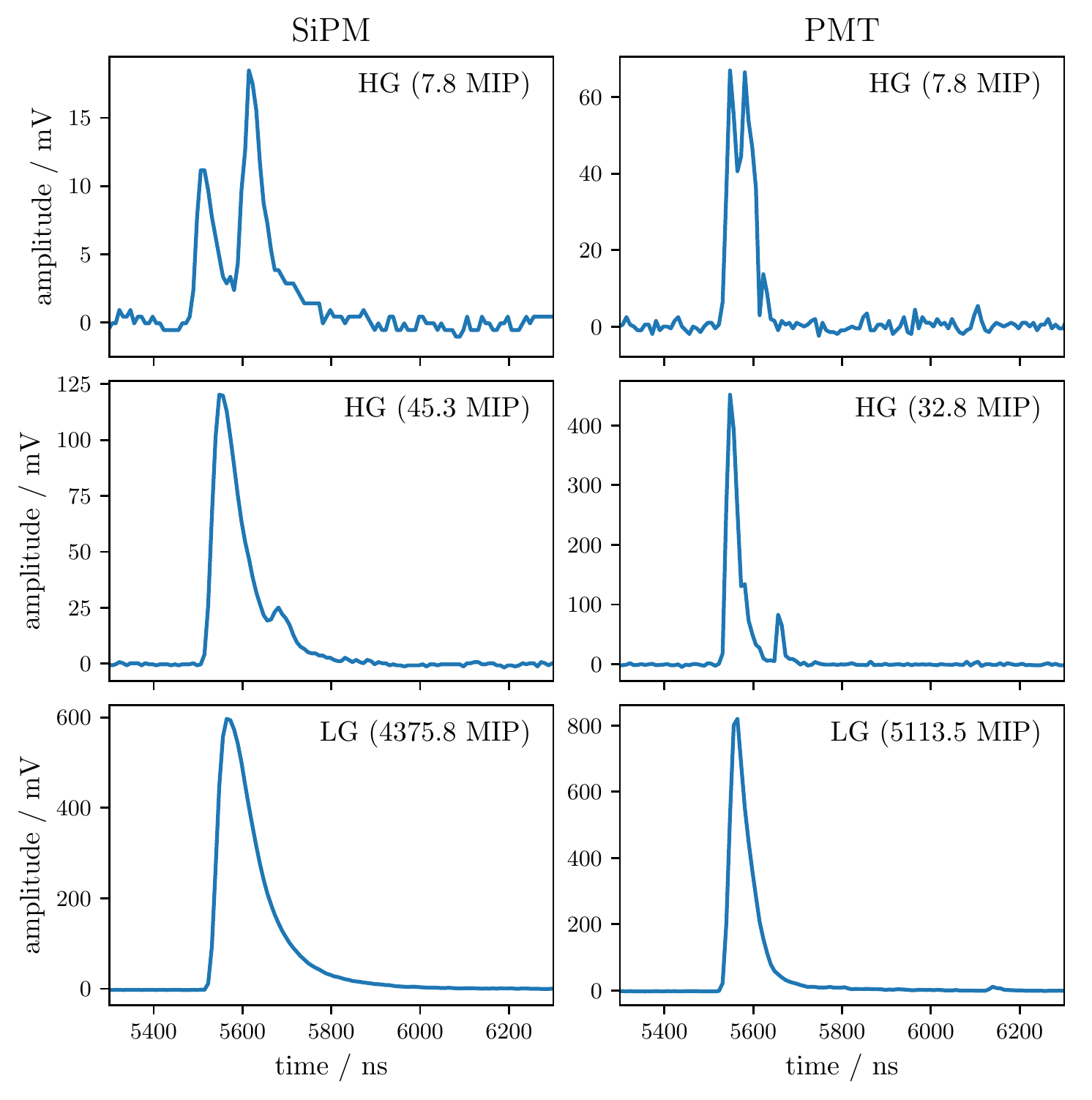}
	\caption{The signal traces measured by the neighbouring SiPM and PMT station for three different events. Note that the detectors are not placed on top of each other. Thus, they do not measure exactly the same signal but only the same shower at slightly different locations.}
	\label{fig:EventTraceExample}
\end{figure}
\begin{figure}\centering
	\includegraphics[]{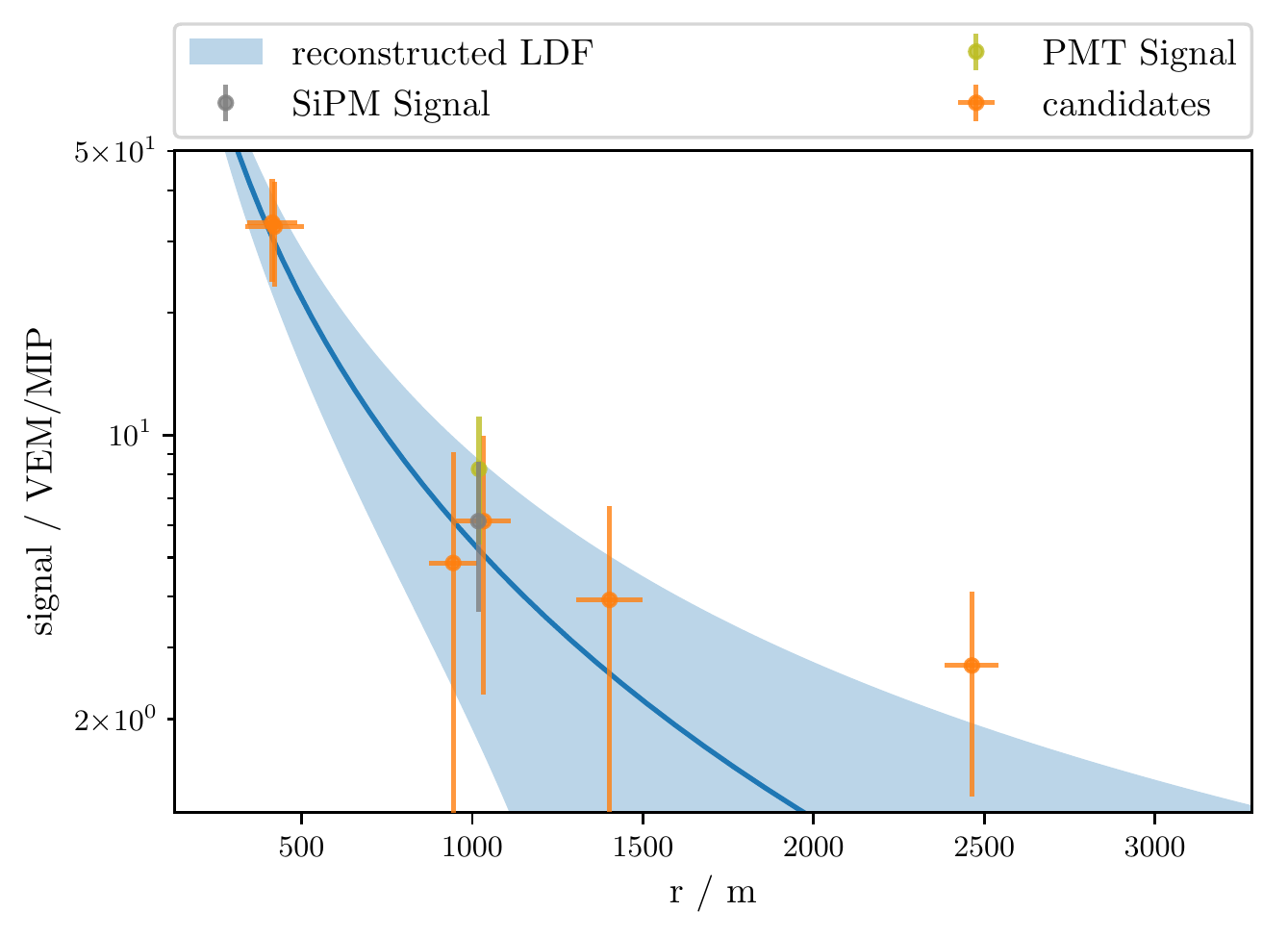}
    \caption{The LDF reconstructed from the signals measured in the surface detector stations. The blue band gives the 1-$\upsigma$ uncertainty region of the reconstruction. The corresponding signal traces for the PMT and SiPM, respectively, are shown in the top row of figure~\ref{fig:EventTraceExample}. The signals in the two SSDs are in agreement with each other and also with the reconstructed LDF.}
    \label{fig:LDFExample}
\end{figure}
Several cosmic-ray air-showers have been detected with the SSD equipped with a SiPM module. Three example traces are shown in figure~\ref{fig:EventTraceExample} and compared to that of the neighbouring detector equipped with a PMT. As the two modules measure the same shower at different locations, the signal peaks are not necessarily correlated. The pulses produced by the SiPM module have a longer decay time than those from the PMT. Due to the precision manufacturing of SiPMs, these pulses have a well-defined shape. It allows for accurate analysis of the pulse structure even when pulses are stacked. 

From the signals of one event measured in coincidence in several surface detector stations, the lateral distribution function (LDF) is reconstructed (see fig.\ \ref{fig:LDFExample}). The LDF describes the expected measured particle density as a function of the distance to the shower core. The signal measured in the SSD is not used in the reconstruction. However, it agrees well with the expectation from the reconstructed LDF. Due to the short time of data taking so far, statistics is very limited but a more sophisticated analysis of the measured events will become possible in the near future.

\section{Summary}
An optical module was developed that exploits the advantages of SiPMs and allows to replace a conventional PMT. A performance at least as good as that of conventional PMTs has been achieved. The optical module is optimized for the application in scintillator-based detectors but also allows for easy usage in laboratory applications as well.

The module features single p.e.\ resolution and a high dynamic range of five orders of magnitude. 
Being powered and controlled via a USB interface it allows for easy usability, in particular in laboratory setups. In addition, the design of the module, including geometry, pulse size and pulse polarity, allows replacing conventional PMTs in many applications.

Its performance has been studied in lab measurements demonstrating a low power consumption below \SI{250}{\mW} and the high dynamic range of \num{e5}. The temperature dependence of the gain of the module is less than \SI{e-4}{\per\kelvin}.

Light guides glued on the SiPMs have been optimized for the read-out of a round bundle of wavelength shifting fibers installed in a scintillator detector to increase the sensitive area. 

The module is installed in an SSD prototype detector for AugerPrime in the Argentinian Pampas. The resulting data proves its usability in harsh environments as well as its performance in astroparticle physics experiments.

The price per module has been estimated to be 530\,\euro{} for a quantity of 10 modules. It reduces to 180\,\euro{} for 1800 modules. In both cases the SiPMs contribute to about \SI{45}{\percent} to the total costs.

\acknowledgments
We acknowledge the financial support of the German Federal Ministry of Education and Research. We further thank the Pierre Auger Collaboration and their local staff for their support and assistance during the installation of the SiPM module and for modifying the scintillator detectors to make them usable with the SiPM module.

\newpage
\bibliographystyle{JHEP}
\bibliography{bibliography}

\end{document}